\documentclass[aps,prb,twocolumn,superscriptaddress]{revtex4-1}
\usepackage{graphicx}
\usepackage{amsmath}
\usepackage{amssymb}
\usepackage{xcolor}

\begin{document}

% Use the \preprint command to place your local institutional report
% number in the upper righthand corner of the title page in preprint mode.
% Multiple \preprint commands are allowed.
% Use the 'preprintnumbers' class option to override journal defaults
% to display numbers if necessary
%\preprint{}

%Title of paper
\title{The magnetic ground state of two isostructual polymeric quantum magnets, [Cu(HF$_{2}$)(pyrazine)$_{2}$]SbF$_{6}$ and [Co(HF$_{2}$)(pyrazine)$_{2}$]SbF$_{6}$, investigated with neutron powder diffraction}

% repeat the \author .. \affiliation  etc. as needed
% \email, \thanks, \homepage, \altaffiliation all apply to the current
% author. Explanatory text should go in the []'s, actual e-mail
% address or url should go in the {}'s for \email and \homepage.
% Please use the appropriate macro foreach each type of information

% \affiliation command applies to all authors since the last
% \affiliation command. The \affiliation command should follow the
% other information
% \affiliation can be followed by \email, \homepage, \thanks as well.
\author{J. Brambleby}\email{J.D.Brambleby@warwick.ac.uk}
\author{P. A. Goddard}
\affiliation{Department of Physics, University of Warwick, Gibbet Hill Road, Coventry, CV4 7AL, UK}
\author{R. D. Johnson}
\affiliation{Department of Physics, Clarendon Laboratory, University of Oxford, Parks Road, Oxford, OX1 3PU, UK}
\author{J. Liu}
\affiliation{Department of Materials,  University of Oxford, 12/13 Parks Road, Oxford, OX1 6PH, UK}
\author{D. Kaminski}
\author{A. Ardavan}
\author{A. J. Steele}
\author{S. J. Blundell}
\affiliation{Department of Physics, Clarendon Laboratory, University of Oxford, Parks Road, Oxford, OX1 3PU, UK}
\author{ T. Lancaster}
\affiliation{Centre for Materials Physics, Durham University, South Road, Durham DH1 3LE, UK}
\author{P. Manuel}
\author{P. J. Baker}
\affiliation{ISIS Pulsed Neutron and Muon Facility, Rutherford-Appleton Laboratory, Chilton, Oxfordshire, OX11 0QX, UK}
\author{J. Singleton}
\affiliation{National High Magnetic Field Laboratory, Los Alamos National Laboratory, MS-E536, Los Alamos, NM 87545, USA}
\author{S. G. Schwalbe}
\author{P. M. Spurgeon}
\author{H. E. Tran}
\author{P. K. Peterson}
\author{J. F. Corbey}
\author{J. L. Manson}\email{jmanson@ewu.edu}
\affiliation{Department of Chemistry and Biochemistry, Eastern Washington University, Cheney, WA 99004, USA}

%\homepage[]{Your web page}
%\thanks{}

%Collaboration name if desired (requires use of superscriptaddress
%option in \documentclass). \noaffiliation is required (may also be
%used with the \author command).
%\collaboration can be followed by \email, \homepage, \thanks as well.
%\collaboration{}
%\noaffiliation

\begin{abstract}
The magnetic ground state of two isostructural coordination polymers (i) the quasi two-dimensional $S$ = 1/2 square-lattice antiferromagnet [Cu(HF$_{2}$)(pyrazine)$_{2}$]SbF$_{6}$; and (ii) a related compound [Co(HF$_{2}$)(pyrazine)$_{2}$]SbF$_{6}$, were examined with neutron powder diffraction measurements.  We find the ordered moments of the Heisenberg $S$  = 1/2 Cu(II) ions in [Cu(HF$_{2}$)(pyrazine)$_{2}$]SbF$_{6}$ are 0.6(1)$\mu_{\textsc{b}}$, whilst the ordered moments for the Co(II) ions in [Co(HF$_{2}$)(pyrazine)$_{2}$]SbF$_{6}$ are 3.02(6)$\mu_{\textsc{b}}$. For Cu(II), this reduced moment indicates the presence of quantum fluctuations below the ordering temperature. We show from heat capacity and electron spin resonance measurements, that due to the crystal electric field splitting of the $S$ = 3/2 Co(II) ions in [Co(HF$_{2}$)(pyrazine)$_{2}$]SbF$_{6}$, this isostructual polymer also behaves as an effective spin-half magnet at low temperatures. The Co moments in [Co(HF$_{2}$)(pyrazine)$_{2}$]SbF$_{6}$ show strong easy-axis anisotropy, neutron diffraction data which do not support the presence of quantum fluctuations in the ground state and heat capacity data which are consistent with 2D or close to 3D spatial exchange anisotropy.
\end{abstract}

% insert suggested PACS numbers in braces on next line
% insert suggested keywords - APS authors don't need to do this
%\keywords{}

%\maketitle must follow title, authors, abstract, \pacs, and \keywords
\maketitle

% body of paper here - Use proper section commands
% References should be done using the \cite, \ref, and \label commands
\section{Introduction}
Low-dimensional magnetic materials exhibit phenomena which cannot be explained semi-classically. One example is quantum fluctuations of magnetic moments at low temperatures, necessitated by the Heisenberg uncertainty principle.\cite{sachdev} In several current theories, quantum fluctuations are predicted to play an important role in quantum phase-transitions in heavy-fermion\cite{hf} and high$-T_{\textsc{c}}$ cuprate superconductors.\cite{sc} Explaining the properties of these condensed-matter systems requires understanding the role of quantum fluctuations across their rich phase-diagrams, which often includes an antiferromagnetic region. In order to explore the macroscopic effects of quantum fluctuations on magnetism alone, it is advantageous to find insulating antiferromagnetic quantum systems which preserve quantum fluctuations to low temperatures.

An example of the macroscopic effects of quantum fluctuations can be seen in low-dimensional antiferromagnets (AFMs). For Heisenberg $S$~=~1/2 moments on a square lattice, quantum-Monte-Carlo simulations predict that quantum fluctuations reduce the ordered moment per ion from its classical value.\cite{SLAFM} Coordination polymers, in which the magnetic exchange between metal ions is mediated by organic molecules, offer the means to test the properties of low-dimensional magnets by constructing three-dimensional lattices of ions with anisotropic exchange pathways. Creating realizations of low-dimensional $S$ = 1/2 AFMs in this way has led to the observation of a reduced magnetic moment for these systems in the ordered phase.\cite{CuClO4, Cu_neutrons}

We use elastic neutron powder diffraction experiments to examine the magnetic ground state of two coordination polymers: (i) [Cu(HF$_{2}$)(pyz)$_{2}$]SbF$_{6}$ (where pyz = pyrazine, C$_{4}$H$_{4}$N$_{2}$), a quasi-two dimensional (Q2D) antiferromagnet based on planes of Heisenberg $S$ = 1/2 moments arranged on a square lattice;\cite{Cu_sat_moment} and (ii) an isostructual polymer [Co(HF$_{2}$)(pyz)$_{2}$]SbF$_{6}$.

\begin{figure}[t]
\includegraphics[width=\columnwidth]{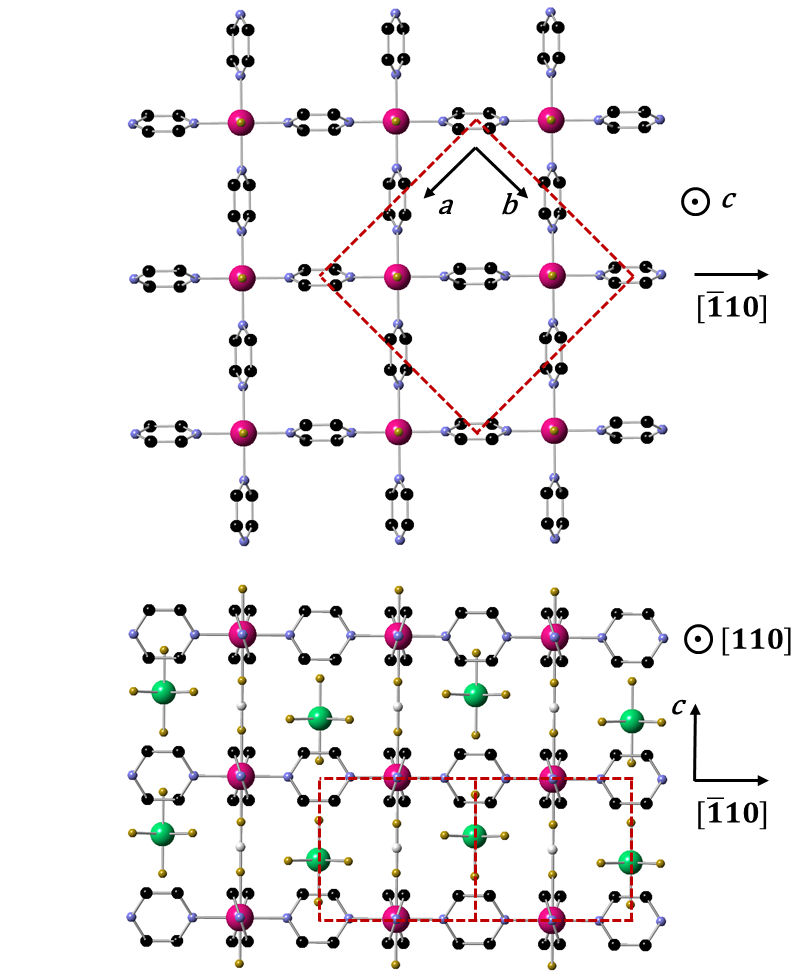}
\caption{Upper image: Structure $[$Cu(HF$_{2}$)(pyz)$_{2}$$]$SbF$_{6}$ viewed along $c$. The SbF$_{6}^{-}$ ions have been omitted for clarity. The Cu(II) ions in $[$Cu(HF$_{2}$)(pyz)$_{2}$$]$SbF$_{6}$ occupy octahedral environments, consisting of two axial fluorine ions from HF$_{2}^{-}$, and four equatorial nitrogen atoms from coordination-bonded pyrazine ligands. The pyrazine ligands form a sheet of co-planar Cu(II) ions, arranged on a square lattice in the $ab$-plane. Lower image:  Structure of $[$Cu(HF$_{2}$)(pyz)$_{2}$$]$SbF$_{6}$ viewed within the $ab-$plane. Adjacent layers are linked by Cu$\cdots$F-H-F$\cdots$Cu pillars along the $c$-axis, and the SbF$_{6}^{-}$ counter-ions occupy the space in the centre of the approximately cubic frameworks.  Cu = pink, C = black, N = blue, F = yellow, Sb = green and H = white (pyrazine hydrogen atoms omitted for clarity) The dashed red line shows the tetragonal structural unit cell boundary. $[$Co(HF$_{2}$)(pyz)$_{2}$$]$SbF$_{6}$ is found to be isotructural, with the Co(II) ions taking the Cu(II) positions in the \textit{P}4/\textit{nmm} unit cell. \label{[Co(HF2)(pyz)2]SbF6_planes}}
\end{figure}

For temperatures in the range 90 to 295~K, [Cu(HF$_{2}$)(pyz)$_{2}$]SbF$_{6}$ is known to crystallise in the \textit{P}4/\textit{nmm} space group. This material consists of planes of Cu(II) ions, each occupying a CuN$_{4}$F$_{2}$ octahedral environment (Fig. \ref{[Co(HF2)(pyz)2]SbF6_planes}).\cite{Cu_sat_moment} Due to a Jahn-Teller distortion along the $c$-axis, the unpaired electron in Cu(II) (3d$^{9}$) occupies the $d_{x^{2}-y^{2}}$ orbital orientated in the $ab$-plane. This leads to significant spin density only along the pyrazine ligand directions, providing the primary antiferromagnetic exchange pathway within the planes ($J_{\parallel}/k_{\textsc{b}}$ = 13.3~K). The material exhibits a transition to long-range order at a finite temperature, $T_{\textsc{n}}$~=~4.3~K, due to the presence of weaker exchange between neighbouring Cu(II) ions in adjacent planes ($J_{\perp}$), where $|J_{\perp}/J_{\parallel}|$ $\simeq$ 9 $\times$ 10$^{-3}$.\cite{Cu_sat_moment, NJP}

For temperatures $T > J_{\parallel}/k_{\textsc{b}}$, the dominant thermal excitations are quasi-independent spin-flips, and for temperatures in the range $T_{\textsc{n}} \le T < J_{\parallel}/k_{\textsc{b}}$, the material behaves as a Q2D magnet, in which regions of antiferromagnetically correlated spins develop in the planes. Renormalization group analysis of the non-linear sigma model \cite{chakravarty1988} suggests that a single 2D plane of interacting spins extracted from a Q2D Heisenberg AFM lies in the renormalized classical regime, where the ground state is N\'{e}el ordered but the moment per ion, $\mu$, is reduced from its classical value, $\mu_{\textsc{c}}$, by quantum fluctuations  at $T$ = 0. In real systems, the finite interplane coupling leads to a regime of long-range magnetic order at $T >$ 0, which can be estimated to occur below a temperature  $T_{\textsc{n}}$, for which

\begin{equation}
J_{\perp}(\mu^{2}/\mu_{\textsc{c}}^{2})(\xi^{2}/a^{2})\approx  k_{\textsc{b}}T_{\textsc{n}},
\label{TN}
\end{equation}
\noindent
where $\xi$ is the correlation length within the planes and $a$ is the lattice constant in the square layers.\cite{chakravarty1988} At low temperatures, we therefore expect the material to exhibit 3D ordering, but with a reduced ordered moment compared to the classical prediction of $\mu/\mu_{\textsc{c}}\approx 0.6$.\cite{SLAFM, chakravarty1988}

Since fluctuations are suppressed by large magnetic fields,\cite{CuClO4} and the bulk magnetization in the ordered phase is zero, we use neutron diffraction as a local zero-field probe of the ordered magnetic structure to test for macroscopic quantum properties. In addition, for this particular material, determining the type of magnetic ordering provides an experimental test of the sign of $J_{\perp}$. The results of the neutron diffraction are compared to those for [Co(HF$_{2}$)(pyz)$_{2}$]SbF$_{6}$. This study provides a means to test how the physics of the Cu(II) material is changed in the related system [Co(HF$_{2}$)(pyz)$_{2}$]SbF$_{6}$ for which each ion site is occupied by Co(II) (3d$^{7}$).  

The Co material was only available as a powder for the experiments described below. For this system, heat capacity measurements show that this material orders at a transition temperature of 7.1~K and we have used neutron powder diffraction to simultaneously solve the nuclear and magnetic structure at 4~K in zero field. We show that this material is isostructural to the Cu sample, and that the magnetic structure is consistent with a local easy-axis anisotropy of each of the Co(II) moments. We report heat capacity and electron spin resonance (ESR) measurements that further support the presence of strong easy-axis anisotropy, which persists up to 30 K and leads to effective spin-half magnetic behaviour below this temperature. The additional effects of anisotropy in this system are explored further with magnetization measurements to map out the field-temperature phase diagram.

\section{Experimental Details}

\textit{Synthesis}: CuF$_{2}$, CoF$_{2}$, and NH$_{4}$HF$_{2}$ were obtained from commercial sources and used as received. Deuterated pyrazine, pyz-D$_{4}$, was purchased from Aldrich Chemical Co. and used without further purification. Only plastic beakers and utensils were used throughout the synthetic process. Because CuF$_{2}$ has low solubility in aqueous solution, 0.4554~g (4.48~mmol) was dissolved in 5-mL HF(aq) (48-52 wt.~$\%$). In a second beaker, the CuF$_{2}$ solution was slowly mixed with a 10-mL aqueous solution containing NH$_{4}$HF$_{2}$ (0.2555~g, 4.48~mmol), NaSbF$_{6}$ (1.1592~g, 4.48~mmol) and pyz-D$_{4}$ (0.7546~g, 8.97~mmol) to afford a deep blue solution. Upon slow evaporation for a few days, a large mass of small deep blue crystals formed. The crystals were collected by suction filtration, washed with 5-mL of fresh H$_{2}$O, and dried \textit{in vacuo} for $\approx$~3 hours, yielding 2.0219~g of [Cu(HF$_{2}$)(pyz-D$_{4}$)$_{2}$]SbF$_{6}$. An analogous synthetic procedure was used to prepare a multi-gram sample of [Co(HF$_{2}$)(pyz-D$_{4}$)$_{2}$]SbF$_{6}$ with CuF$_{2}$ being replaced by CoF$_{2}$. Infrared spectroscopy confirmed the presence of HF$_{2}^{-}$, pyz-D$_{4}$ and SbF$_{6}^{-}$ in each of the two products.

The large single crystal of [Cu(HF$_{2}$)(pyz)$_{2}$]SbF$_{6}$ used in the DC magnetometry study (below) was slowly grown from HF(aq) solution using hydrogenated pyrazine, pyz-H$_{4}$. Small deep blue crystals were allowed to grow for 2 days and removed from solution. The mother liquor was left to stand undisturbed at room temperature for 1 month.

\textit{Neutron Powder Diffraction}: Several grams of each sample were produced in which pyrazine was replaced with deuterated pyrazine (pyz-D$_{4}$ = C$_{4}$D$_{4}$N$_{2}$) at the sample synthesis stage. This reduces the incoherent scattering of neutrons from hydrogen in the sample, lessening the extent of the neutron beam attenuation and increasing the signal-to-noise ratio for the weak magnetic scattering. The polycrystalline samples were ground into a fine powder, loaded into a vanadium can and zero-field elastic neutron scattering experiments were performed using WISH at ISIS, Rutherford Appleton Laboratory. \cite{WISH} Diffraction patterns were collected at temperatures: (i) 1.5~K and 5~K for [Cu(HF$_{2}$)(pyz-D$_{4}$)$_{2}$]SbF$_{6}$; and (ii) 4~K and 10~K for [Co(HF$_{2}$)(pyz-D$_{4}$)$_{2}$]SbF$_{6}$. These were chosen to be above and below the respective ordering temperatures.

\textit{Heat Capacity}: Heat capacity measurements were performed on a 4.4(1)~mg polycrystalline sample of [Co(HF$_{2}$)(pyz)$_{2}$]SbF$_{6}$ with a Quantum Design PPMS. The sample was held in thermal equilibrium with a large thermal reservoir, and the temperature of the stage/sample was monitored as a heat pulse was applied to the system. The heat capacity is then extracted by fitting the time-dependent response.\cite{Lashley} The data were corrected for the heat capacity of the stage/grease by subtracting a measurement of the heat capacity when no sample was present. The field-dependence of the magnetic phase transition was investigated in applied fields up to $\mu_{0}H = 12$ T.

\textit{ESR}: High frequency electron-spin resonance measurements were performed on a finely ground sample of [Co(HF$_{2}$)(pyz)$_{2}$]SbF$_{6}$ using the 15 T homodyne transmission ESR spectrometer at the National High Magnetic Field Laboratory, Tallahassee. The sample was loosely restrained in a Teflon container. A phase-locked oscillator, in conjunction with a series of multipliers and amplifiers, was employed as a microwave source capable of providing quasi-continuous frequency coverage up to 600~GHz, and a cold bolometer was used for detection.

\textit{DC Magnetometry}: The magnetic moment ($M$)  of [Co(HF$_{2}$(pyz)$_{2}$]SbF$_{6}$ was measured as a function of applied field ($H$) with an Oxford Instruments Vibrating Sample Magnetometer (VSM). A powdered sample was pressed into a pellet, attached to a PEEK rod with vacuum grease, and wrapped in PTFE tape to prevent the sample from moving. Data were recorded on sweeping the field from 0 $\le \mu_{0}H \le$ 12 T, for fixed temperatures in the range 1.4~$\le T \le 8$ K.

The temperature dependence of ($M$) was recorded with a VSM technique for 4 $\le T \le$ 305~K in a constant field $\mu_{0}H$ = 0.1~T.

An investigation of the anisotropic magnetic properties of a 17.9 mg single crystal of [Cu(HF$_{2}$)(pyz)$_{2}$]SbF$_{6}$ was performed with a Quantum Design SQUID Magnetometer. The crystal grew as a square plate, and the $\textbf{c}$-axis was taken perpendicular to this plane. The sample was mounted in a gelatin capsule with cotton wool to hold it in place, and fixed inside a plastic drinking straw. The sample's magnetic moment was then measured, as a function of field (at 2 K) and temperature (in a DC-field $\mu_{0}H = 0.1$ T), for fields $\parallel \textit{\textbf{c}}$ and $\perp \textit{\textbf{c}}$.  In the linear limit, the molar susceptibility ($\chi_{\text{mol}}$) is obtained from this measurement using the relation $\chi_{\text{mol}} = M/nH$, where $n$ is the number of moles of the sample.

\textit{Pulsed Field Magnetometry}:  The isothermal pulsed-field magnetization, in fields up to $\mu_{0}H$ = 20 T, was measured with a capacitor-bank powered, short-pulse magnet, with a rise time to maximum field $\approx10$~ms. A polycrystalline sample was packed into a PCTFE ampoule of diameter 1.3 mm and sealed with vacuum grease to prevent the sample from moving. The filled ampoule was lowered into a 1.5 mm bore, 1500-turn compensated-coil susceptometer, which is 1.5 mm in length and made from high-purity copper wire. With the sample at the centre of the coil, the voltage induced in the coil as the field is pulsed is proportional to the rate of change of magnetization. After subtracting the signal measured from an empty coil, under the same thermal conditions, and numerically integrating the difference with respect to time, the magnetization of the sample can be determined. The value of the field is deduced by measuring the de Haas-van Alphen oscillations induced in the measurement coil.

\begin{table*}
 \caption{Refinement details from neutron powder diffraction data for \textbf{(a)} [Cu(HF$_{2}$)(pyz-D$_{4}$)$_{2}$]SbF$_{6}$ at 1.5 K; and \textbf{(b)} [Co(HF$_{2}$)(pyz-D$_{4}$)$_{2}$]SbF$_{6}$, at 4 K. $Z$ is the number of formula units per unit cell; \textbf{\textit{k}} is the magnetic propagation vector in reciprocal lattice units (r.l.u.); $\mu$ is the refined magnetic moment of \textbf{(a)} Cu(II) and \textbf{(b)} Co(II). The goodness-of-fit parameters $R_\text{F}$ = $\sum|I_{\text{obs}}^{1/2}-I_{\text{calc}}^{1/2}|/\sum I_{\text{obs}}^{1/2}$ and $R_\text{Bragg}$ =  $\sum |I_{\text{obs}}-I_{\text{calc}}|/\sum|I_\text{obs}|$ where $I_\text{obs}$ is the observed intensity, $I_\text{calc}$ is the calculated intensity and the sum runs over all data points. $ R_\text{mag}$ is the equivalent $R$-factor to $R_{\text{Bragg}}$ applied to the fit of the magnetic scattering only. \label{Cu_refinement}}
\begin{ruledtabular}
 \begin{tabular}{l c l l}
Parameter (Units) & \textbf{(a)} Fit Result (Error) & \multicolumn{2}{c}{\textbf{(b)} Fit Result (Error)}\\
\hline
Material & [Cu(HF$_{2}$)(pyz-D$_{4}$)$_{2}$]SbF$_{6}$  & \multicolumn{2}{c}{[Co(HF$_{2}$)(pyz-D$_{4}$)$_{2}$]SbF$_{6}$}\\
Formula & C$_{8}$HD$_{8}$N$_{4}$F$_{8}$CuSb & \multicolumn{2}{c}{C$_{8}$HD$_{8}$N$_{4}$F$_{8}$CoSb} \\
Temperature (K) & 1.5 & \multicolumn{2}{c}{4} \\
Space Group & \textit{P}4/\textit{nmm}  & \multicolumn{2}{c}{\textit{P}4/\textit{nmm}}\\
$a$ (\text{\AA}) & 9.6749(1) & \multicolumn{2}{c}{10.0225(8)}  \\
$c$ (\text{\AA}) & 6.7981(3) & \multicolumn{2}{c}{ 6.4287(5)} \\
$Z$ & 2 & \multicolumn{2}{c}{2} \\
Unit Cell Vol. (\text{\AA}$^{3}$) & 636.33(3) & \multicolumn{2}{c}{645.76(9)}\\
\textbf{\textit{k}} (r.l.u.)& (0, 0, 1/2) &  \multicolumn{2}{c}{(0, 0, 1/2)}\\
Order Type & G-Type & \multicolumn{2}{c}{G-Type}\\
$\mu$ ($\mu_{\textsc{b}}$) & 0.6(1) & \multicolumn{2}{c}{3.02(6)}\\
\hline
Detector bank & Bank 2 & Bank 2 & Bank 5\\
\hline
$R_{\text{F}}$ & 0.0944 & 0.0369 &  0.0433 \\
$R_{\text{Bragg}}$ & 0.0256 & 0.0402 & 0.0764 \\
$R_{\text{mag}}$ & 0.170\textcolor{white}{0} & 0.050 & 0.134 \\
\end{tabular}
\end{ruledtabular}
\end{table*}

\section{Results}
\subsection{ [Cu(HF$_{2}$)(pyz)$_{2}$]SbF$_{6}$}
\subsubsection{Neutron Powder Diffraction}

\begin{figure}[b]
\includegraphics[width=\columnwidth]{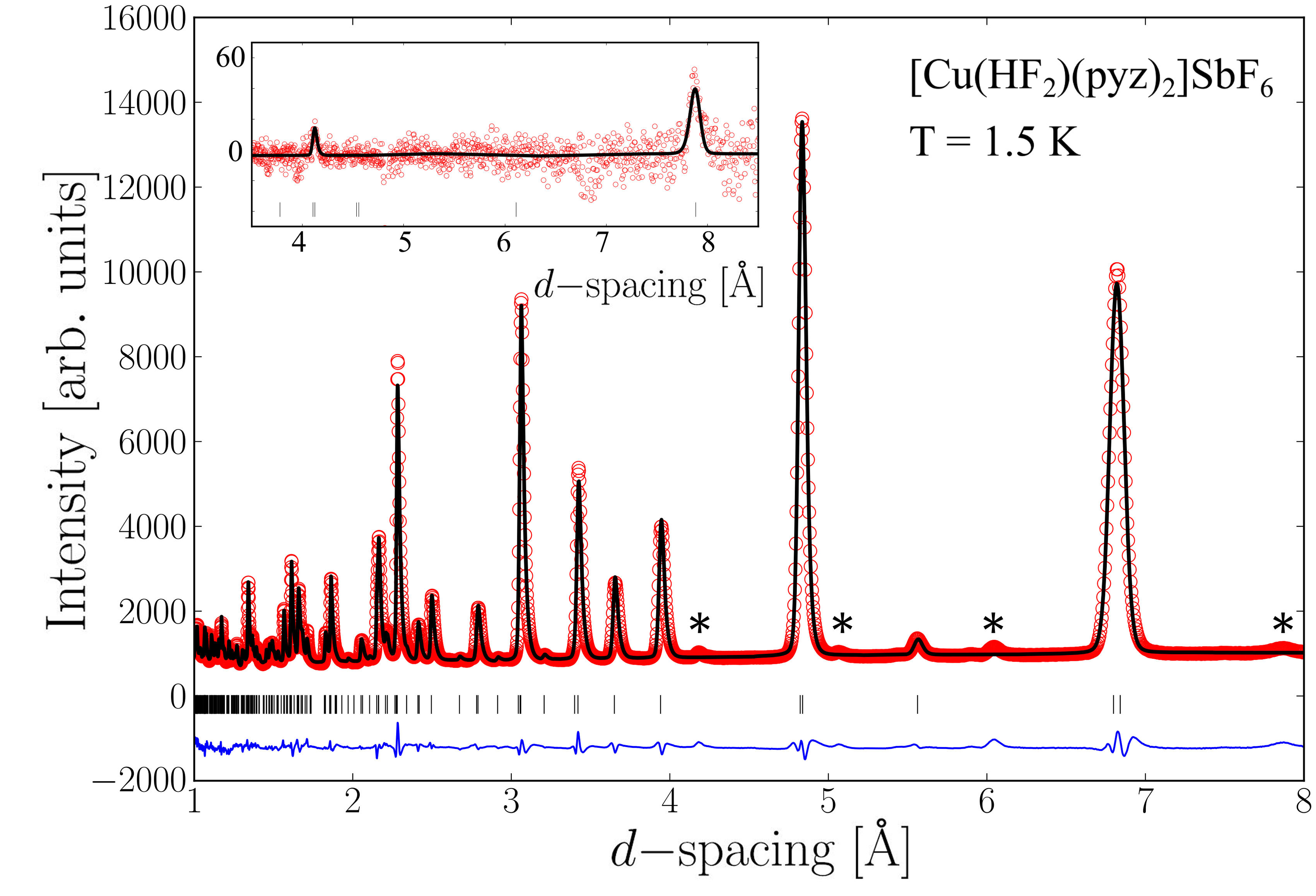}
\caption{Refinement of neutron diffraction data from a powdered sample of $[$Cu(HF$_{2}$)(pyz-D$_{4}$)$_{2}$$]$SbF$_{6}$ at 1.5 K. Main Panel: Bank 2 data ($I_{\text{obs}}$, red circles), fitted model ($I_{\text{calc}}$, black line), reflections of the \textit{P}4/\textit{nmm} lattice (black ticks), and $I_{\text{obs}}-I_{\text{calc}}$ (blue line). Impurity peaks are marked by an asterix, but do not affect the refinement. If the impurity is a copper based polymeric compound, the intensity of these  impurity peaks imply it is $< 1\%$ of the sample and was not detectable in the magnetometry measurements. Inset: Magnetic peaks extracted from the data (see text). The $y$-axis of the inset is in the same units as that of the main panel.\label{CuSbF6_neutrons}}
\end{figure}

A plot of the intensity of neutrons scattered from [Cu(HF$_{2}$)(pyz-D$_{4}$)$_{2}$]SbF$_{6}$ (collected in Bank 2 of the instrument WISH) as a function of the plane spacing ($d$) of the nuclei/moments from which the neutrons are scattered at 1.5 K is shown in Fig. \ref{CuSbF6_neutrons}. By subtracting data collected at 5 K from the 1.5 K data, we observe the development of magnetic Bragg peaks, present below the ordering temperature $T_{\textsc{n}}$ = 4.3 K (Fig. \ref{CuSbF6_neutrons}, inset). These peaks are evident at large $d$-spacings only, which is consistent with the magnetic form factor of Cu(II)\cite{form} supporting the conclusion that they are magnetic in origin.

The published\cite{Cu_sat_moment} high temperature structure of [Cu(HF$_{2}$)(pyz)$_{2}$]SbF$_{6}$ was refined using \textsc{fullprof}\cite{fullprof} against nuclear diffraction peaks measured at 1.5 K.  In order for the crystal structure refinement to reach convergence, it was necessary to fix the hydrogen isotropic thermal parameters to a constant, while those for the independent fluorine ions in the unit cell were coupled to each other. Each hydrogen from pyrazine in the published structure was also replaced with the deuterium isotope. A summary of the refinement details is given in Table \ref{Cu_refinement} and a full list of the refined atomic positions can be found in the crystallographic information file (CIF) attached as supplementary information.\cite{supp}

We find that the crystal structure at 1.5 K retains its tetragonal unit cell, with the space group \textit{P}4/\textit{nmm} (Fig.~\ref{[Co(HF2)(pyz)2]SbF6_planes}), implying that there are no structural phase transitions in this compound on cooling from room temperature to 1.5~K.\cite{Cu_sat_moment} The lattice parameters were found to be $a = b$ = 9.6748(1)~\text{\AA} and $c$ = 6.7988(3)~\text{\AA}, which are slightly reduced from those published for the fully hydrogenated structure at 90 K.\cite{Cu_sat_moment}  The structural refinement demonstrates that the deuteration of pyrazine has little impact on the crystal structure. We therefore expect the magnetic properties of this sample to be similar to those of the fully hydrogenated sample. 

The magnetic peaks observed below $T_{\textsc{n}}$ could be indexed with the commensurate propagation vector \textit{\textbf{k}} = (0, 0, 1/2) (in r.l.u.), indicating a doubling of the unit cell along the $c$-axis. Since there are two coplanar Cu(II) ions in the structural unit cell (Fig.~\ref{[Co(HF2)(pyz)2]SbF6_planes}), the magnetic propagation vector indicates there are four Cu(II) ions in the magnetic super-cell. The full magnetic structure can then be expressed by determining the orientation of these four moments.

The doubling of the unit cell along $c$ indicates that magnetic coupling of Cu(II) ions in adjacent planes must be antiferromagnetic, restricting the free parameters of the magnetic structure to the orientation of the two coplanar Cu(II) ions in the super-cell. Symmetry analysis limits the allowed relative orientations of the two moments to four cases. The moments must be collinear, ferromagnetically or antiferromagnetically (AFM) aligned, with directions that lie (i) exactly along $c$; or (ii) in an arbitrary direction in the $ab-$plane.

By comparing the position of the measured magnetic Bragg peaks (Fig.~\ref{CuSbF6_neutrons}) to those expected from each of the four cases above, we find that only models in which the coplanar Cu(II) moments are AFM coupled can generate magnetic scattering at the peak positions observed in the neutron scattering measurement. We can therefore robustly determine that the moments are AFM coupled in all directions. This is G-type AFM order.\cite{Blundell} 

We tested whether the relative intensity of the magnetic Bragg peaks could be used to differentiate between cases (i) and (ii) within the G-type model. However, due to the weak magnetic reflections observed, the difference in relative intensity of magnetic peaks expected from the two cases is too small for the measured data to be used to determine the Cu(II) moment direction. 

Having established the G-type model and given the result that, within this model, the relative intensity of the peaks is insensitive to the moment direction, the absolute intensity of the magnetic diffraction pattern (scaled to the nuclear reflections) can be used to determine the size of the Cu(II) moments. The nuclear structure, the Cu(II) magnetic moment and associated errors were refined simultaneously, with the magnetic scattering refined only against the subtracted data [Fig. \ref{CuSbF6_neutrons}(inset)]. We find that the size of the moment for each $S$ = 1/2 Cu(II) ion is $\mu_{\text{Cu}}$~=~0.6(1)$\mu_{\textsc{b}}$, where the error considers the covariance parameters of the magnetic model under consideration. This measurement of the reduced ordered moment shows that [Cu(HF$_{2})$(pyz)$_{2}$]SbF$_{6}$ exhibits quantum fluctuations which are present at the finite temperature of 1.5 K, well below the ordering temperature of 4.3 K.

\subsubsection{Single Crystal DC Magnetometry}

\begin{figure}[t]
\includegraphics[width=\columnwidth]{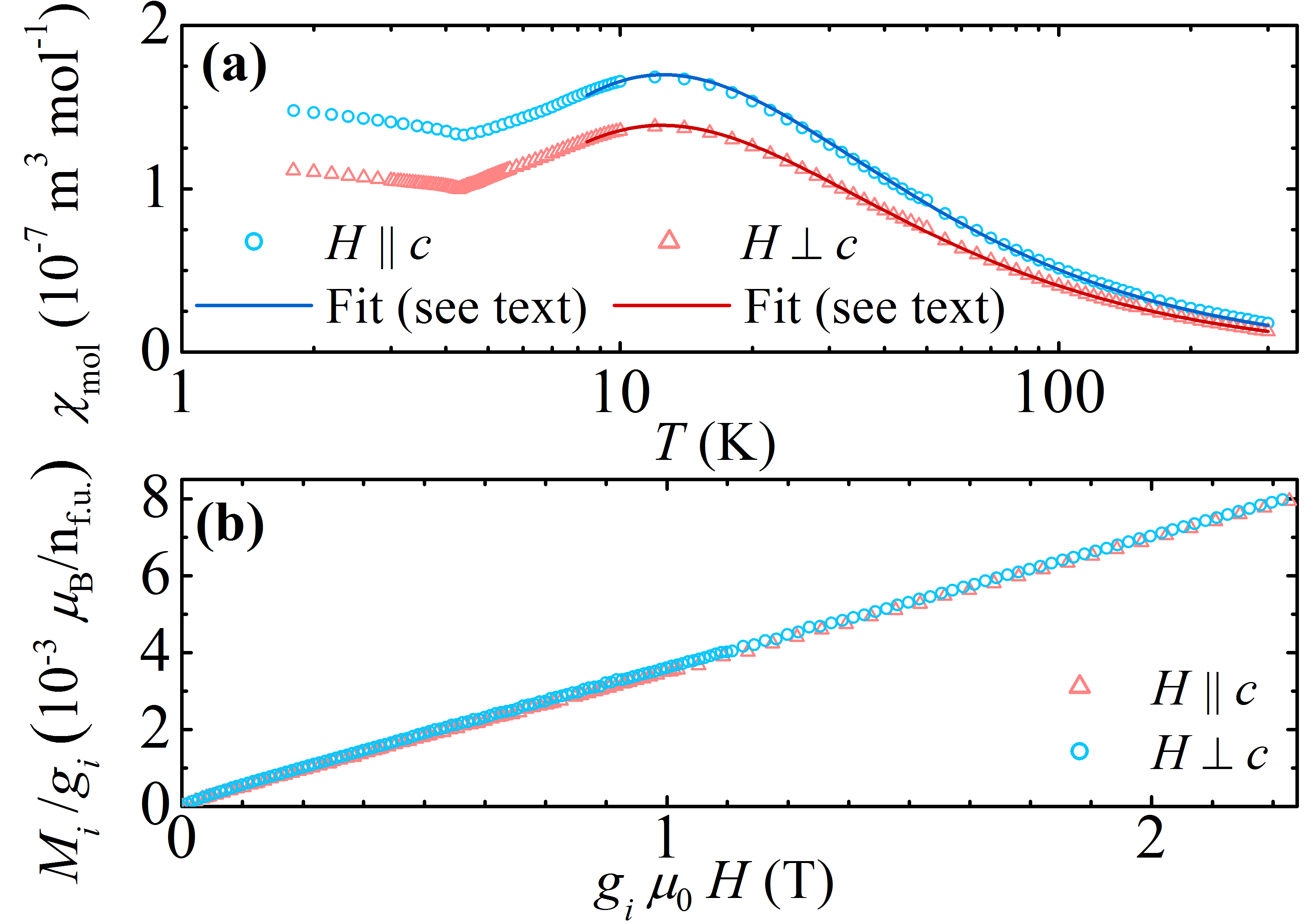}
\caption{ \textbf{(a)} Susceptibility vs. temperature measured in an applied field $\mu_{0}H$ = 0.1 T for an orientated single crystal of [Cu(HF$_{2}$)(pyz)$_{2}$]SbF$_{6}$. Data have been fitted to a model of $S$ = 1/2 moments on a square lattice with nearest neighbour interactions. The fit yields $g_{xy}$ = 2.06(2), $g_{z}$ = 2.28(3) and $J_{\parallel}/k_{\textsc{b}}$ = 13.3(1) K. For $H \perp c$, the fitted values of the parameters $J$ and $g$ did not change within the reported uncertainty when the tetragonal system was rotated by 45$^{\circ}$ about $c$. \textbf{(b)} Magnetization at $T$ = 2 K (in units of Bohr magnetons per formula unit, $n_{\text{f.u.}}$) normalised by the relevant $g$-factor ($g_{i}$ where $i$ = z, xy) vs. applied field scaled by the same $g$-factor.\label{CuSbF6_mvsB}}
\end{figure}

The DC magnetic properties of a single crystal sample of [Cu(HF$_{2}$)(pyz)$_{2}$]SbF$_{6}$ were studied to try to determine the orientation of the ordered moments below $T_{\textsc{n}}$. The molar susceptibility ($\chi_{\text{mol}}$) vs. temperature, for applied fields $H \parallel \textit{\textbf{c}}$ and $H \perp \textit{\textbf{c}}$, is shown in Fig. \ref{CuSbF6_mvsB}(a). The data from 8 $\le T \le$ 300 K were fitted to a model of Heisenberg $S$ = 1/2 moments arranged on a square lattice with antiferromagnetic nearest neighbour exchange, $J_{\parallel}$.\cite{woodward1} The small diamagnetism of the sample holder was accounted for with a temperature-independent correction ($\chi_{0}$ = $-$3.2 $\times$ 10$^{-9}$ m$^{3}$ mol$^{-1}$).

The fits show the Cu(II) ions have a $g$-factor anisotropy, which arises due to spin-orbit coupling.\cite{Kahn} For  $H \parallel \textit{\textbf{c}}$, the fitting yields $g_{z}$ = 2.28(3), whilst for  $H \perp \textit{\textbf{c}}$, $g_{xy}$ = 2.06(2). A powder average $g$-factor of these results, 2.14(3), agrees with the published value.\cite{NJP} In addition, the exchange parameter determined from both fits, $J_{\parallel}/k_{\textsc{b}}$ = 13.3(1) K, agrees with previous results.\cite{Cu_sat_moment, NJP} Using the same two orientations of the crystal, the magnetization at 2 K was measured in an applied field, shown in Fig.~\ref{CuSbF6_mvsB}(b). The axes were scaled to remove the effect of the $g$-factor anisotropy on the size of $M$. We find that the rescaled-magnetization is almost identical for fields $\parallel$ and $\perp$ to $\textit{\textbf{c}}$, indicating, that at these temperatures, there was no measurable anisotropy of the Cu(II) ions beyond spin-orbit coupling regularly observed in these compounds.\cite{NJP}

\subsection{$[$Co(HF$_{2}$)(pyz)$_{2}$$]$SbF$_{6}$}
\subsubsection{Neutron Powder Diffraction}

\begin{figure}[t]
\includegraphics[width=\columnwidth]{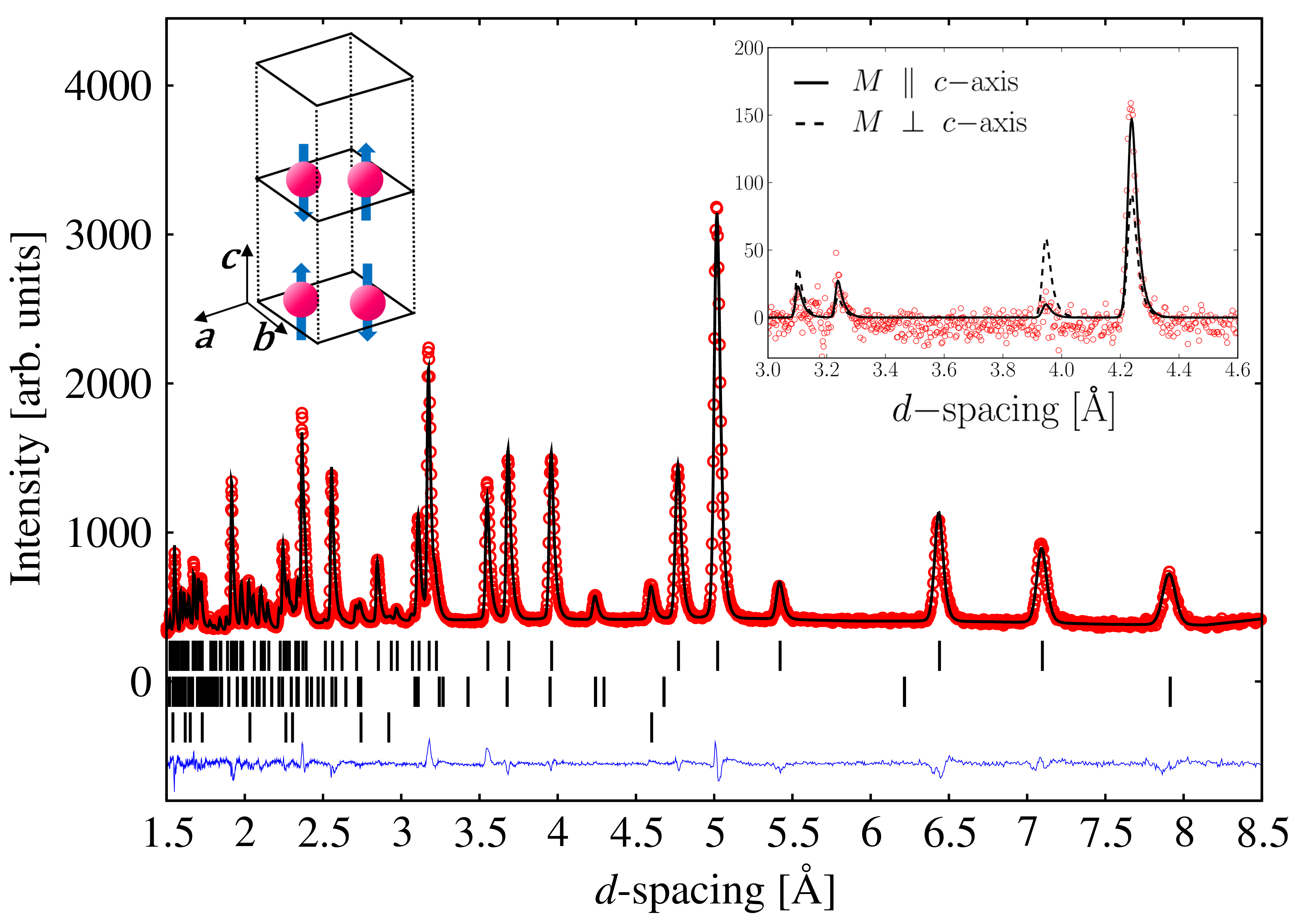}
\caption{Main Panel: Refinement of powder neutron-diffraction data for [Co(HF$_{2}$)(pyz-D$_{4}$)$_{2}$]SbF$_{6}$ at 4 K. Bank 2 data (red circles), fitted model (black line), reflections of the P4/\textit{nmm} lattice (black ticks, top line), reflections of the magnetic moments with propagation vector \textit{\textbf{k}} = (0, 0, 1/2) (black ticks, bottom line) and $I_{\text{obs}}-I_{\text{calc}}$ (blue line). The third row of ticks indicate structural peaks of at 12.4(7)$\%$ non-magnetic crystalline impurity of NaHF$_{2}$ ($R\bar{3}m$), likely formed in the synthesis of [Co(HF$_{2}$)(pyz)$_{2}$]SbF$_{6}$ from a reaction of two precursor compounds NaSbF$_{6}$ and NH$_{4}$HF$_{2}$. Inset (right): Magnetic peaks extracted from the data by subtracting a diffraction pattern collected at 10 K from the 4 K data set (the units of the $y$-axis are the same of those of the main panel). The lines represent the expected magnetic peaks when the moments are aligned parallel (solid line) and perpendicular (dashed line) to the $c$-axis. Inset (left): Schematic of the proposed magnetic supercell of the ground state. Pink = cobalt ion, blue arrow = ordered moment direction. All non-magnetic ions have been omitted for clarity. \label{CoSbF6_neutrons}}
\end{figure}

Zero-field neutron powder diffraction data for [Co(HF$_{2}$)(pyz-D$_{4}$)$_{2}$]SbF$_{6}$ were collected using the instrument WISH. Data from two banks of detectors were used in the analysis, these were (i) Bank 2, a group of detectors at low scattering angles with a high flux at long $d$-spacings suited for looking at magnetic Bragg peaks; and (ii) Bank 5, which is located at large scattering angles for high structural resolution. A plot of the diffraction data (collected in Bank 2, with the sample at $T$ = 4 K) is shown in Fig. \ref{CoSbF6_neutrons}. The difference in scattering between data collected at 4 K and 10 K is shown in the inset to Fig. \ref{CoSbF6_neutrons}. The peaks which remain after this subtraction occur at $d$-spacings which do not correspond to nuclear Bragg peaks, which suggests they originate from antiferromagnetically ordered Co(II) moments.

\begin{figure}[b]
\includegraphics[width=\columnwidth]{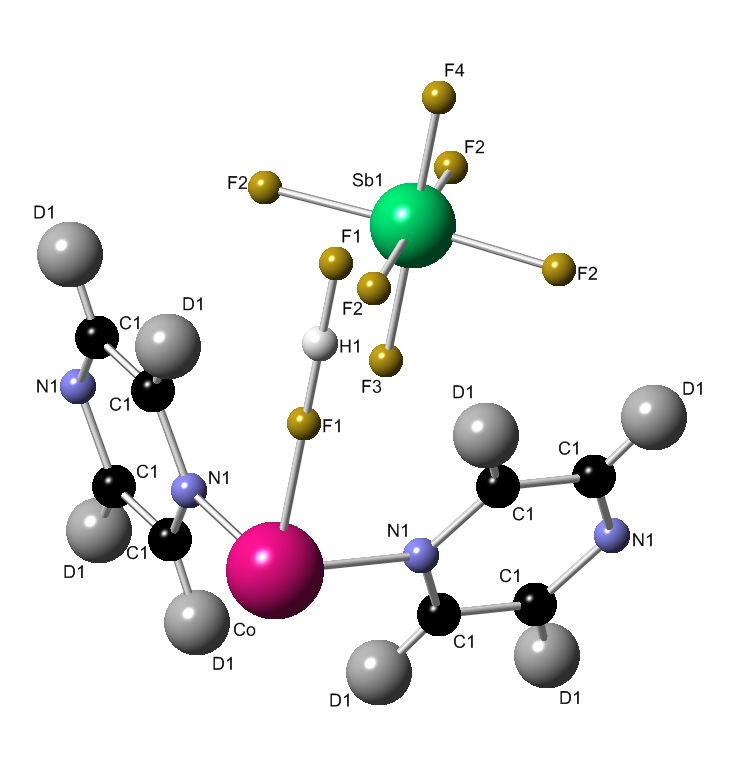}
\caption{Atom labelling scheme for the refined atomic positions in the formula unit [Co(HF$_{2}$)(pyz-D$_{4}$)$_{2}$]SbF$_{6}$. Note: site D1 is refined with the deuterium isotope (grey), whereas site H1 is refined with the hydrogen isotope (white). Duplicated labels are equivalent positions in the \textit{P}4/\textit{nmm} space group.\label{[Co(HF2)(pyz)2]SbF6_formula_unit}}
\end{figure}

Starting from the published\cite{Cu_sat_moment} structure of $[$Cu(HF$_{2}$)(pyz)$_{2}$$]$SbF$_{6}$, the structure of $[$Co(HF$_{2}$)(pyz-D$_{4}$)$_{2}$$]$SbF$_{6}$ was refined using \textsc{fullprof} against the nuclear Bragg peaks collected at 4 K in both detector banks. We find $[$Co(HF$_{2}$)(pyz)$_{2}$$]$SbF$_{6}$ is isostructural to $[$Cu(HF$_{2}$)(pyz)$_{2}$$]$SbF$_{6}$, crystallising in the space group \textit{P}4/\textit{nmm} (Fig. \ref{[Co(HF2)(pyz)2]SbF6_planes}). A summary of the refinement details, compared to those for the Cu sample, are given in Table~\ref{Cu_refinement}. A full list of the refined atomic positions is given in Table \ref{Co_atoms}, and the labelling scheme for each atom in the unit cell is given in Fig. \ref{[Co(HF2)(pyz)2]SbF6_formula_unit}. A CIF file of the 4 K structure is attached as supplementary information.\cite{supp}

The Co(II) ions occupy sites in the tetragonal unit cell with point group symmetry $\bar{4}m2$. This restricts the axes of the CoN$_{4}$F$_{2}$ to maintain the orientation shown in Fig.~\ref{[Co(HF2)(pyz)2]SbF6_planes}, but permits a distortion such that the 4 equidistant Co-N bond lengths ($x$), and the 2 equidistant Co-F bond lengths ($y$) may differ from each other. This octahedral distortion can be parametrised as $y$/$x$, and using the measured bond distances at 4 K, $x$ = 2.146~\text{\AA} and $y$ = 2.091~\text{\AA} (Table~\ref{[Co(HF2)(pyz)2]SbF6_formula_unit}), we find $y$/$x$ = 0.97 such that the octahedra are axially compressed. This is in contrast to the Cu sample where a Jahn-Teller distortion of the CuN$_{4}$F$_{2}$ octahedra gives rise to the measured distortion $y$/$x$ = 1.08 at 1.5 K.

\begin{table}[t]
 \caption{Refined fractional atomic coordinates for [Co(HF$_{2}$)(pyz-D$_{4}$)$_{2}$]SbF$_{6}$ at 4 K. Positions without errors are fixed by the symmetry of the \textit{P}4/\textit{nmm} space-group. The labelling scheme for atoms is shown in Fig. \ref{[Co(HF2)(pyz)2]SbF6_formula_unit}.\label{Co_atoms}}
\begin{ruledtabular}
 \begin{tabular}{l l l l}
Atom & \multicolumn{3}{c}{Fractional coordinates}\\
 & $x$ & $y$ & \textcolor{white}{-}$c$\\
\hline
Co & 0.7500 & 0.2500 & \textcolor{white}{-}0.0000\\
F1 & 0.7500 & 0.2500 & -0.325(1)\\
H1 & 0.7500 & 0.2500 & \textcolor{white}{-}0.5000\\
N1 & 0.5986(2)& 0.0986(2) & \textcolor{white}{-}0.0000 \\
C1 & 0.5745(2)& 0.0245(2) & \textcolor{white}{-}0.1693(5)\\
D1 & 0.6315(3)& 0.0430(3) & \textcolor{white}{-}0.3121(6)\\
Sb1 & 0.2500 & 0.2500 & \textcolor{white}{-}0.410(1) \\
F2 & 0.4417(4)& 0.2500 & \textcolor{white}{-}0.4062(8)\\
F3 & 0.2500 & 0.2500 & \textcolor{white}{-}0.114(2) \\
F4 & 0.2500 & 0.2500 & \textcolor{white}{-}0.693(2) \\
\end{tabular}
\end{ruledtabular}
\end{table}

The lattice parameters were found to be $a$~=~$b$~=~10.0225(8)~\text{\AA} and $c$~=~6.4287(5)~\text{\AA}, indicating the [Co(pyz)$_{2}$]$^{2+}$ sheets are more closely stacked than in the Cu sample. The pyrazine molecules in the Co sample are also rotated further from the vertical, with a tilt angle of $\pm$18.03(4)$^{\circ}$ compared to $\pm$12.2(5)$^{\circ}$ for the 1.5~K structure of the Cu sample.

From the neutron powder diffraction below $T_{\textsc{n}}$, the observed magnetic Bragg peaks (Fig. \ref{CoSbF6_neutrons} inset)  were indexed with a commensurate propagation vector \textbf{\textit{k}} = $\left(0, 0, 1/2\right)$ (in r.l.u.) again indicating a doubling of the unit-cell along the $c$-axis. By analogy with the Cu sample, this results in four Co(II) ions in the magnetic super-cell. 

The analysis of the magnetic neutron diffraction proceeds in the same way as outlined in Section III.A. The propagation vector implies that Co(II) ions in adjacent planes are antiferromagnetically aligned, and the observed position of the magnetic reflections is only consistent with AFM coupling between the coplanar Co(II) ions. This allowed the unambiguous determination of G-type ordering in the ground state.

The relative intensities of the magnetic Bragg peaks predicted by each of the two possible groundstates were compared to the measured diffraction data, (Fig. \ref{CoSbF6_neutrons}, inset). A model in which the moments are parallel to $\textbf{c}$ provides the best representation of the experimental data, indicating the moments in the groundstate exhibit local easy-axis anisotropy.

Within the G-type model with the Co moments parallel to the $c$-axis, the size of the ordered moment ($\mu_{\text{Co}}$) could be refined simultaneously against the diffraction data along with the nuclear structure. Using the data collected in both Bank 2 and Bank 5,  we find $\mu_{\text{Co}}$~=~3.02(6)$\mu_{\textsc{b}}$. This is consistent with the moment of Co(II) ions in the high-spin state ($S$ = 3/2).

\begin{figure}[t]
\includegraphics[width=\columnwidth]{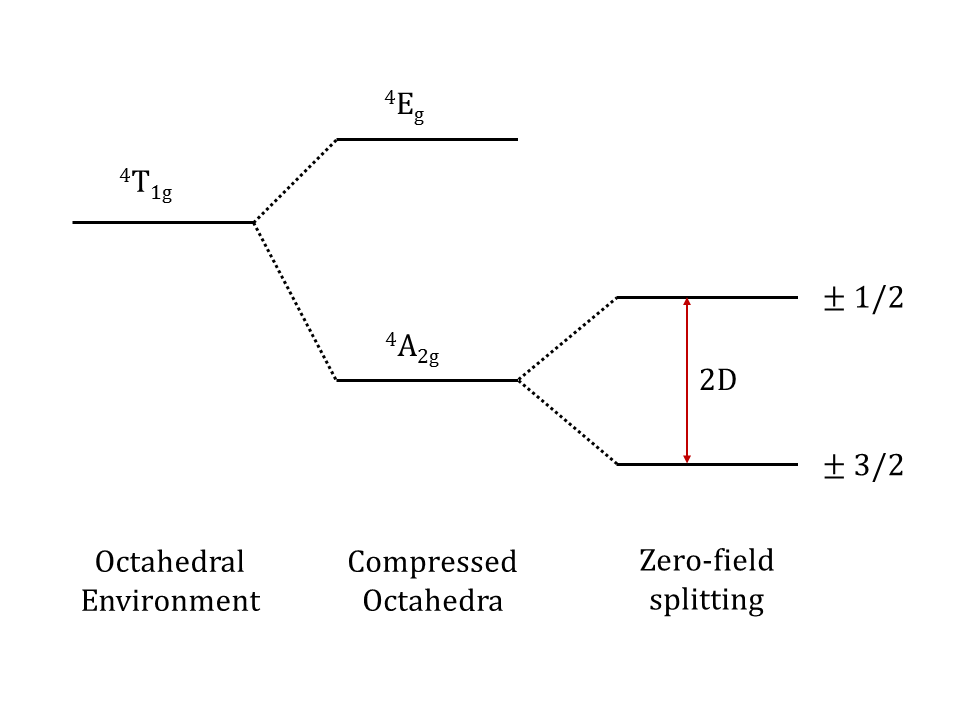}
\caption{Energy level diagram for high-spin Co(II) ions in an octahedral environment,\cite{Co2} applied to [Co(HF$_{2})$(pyz)$_{2}$]SbF$_{6}$. The axial compression of the CoN$_{4}$F$_{2}$ octahedra along the Co$-$F bond lifts the degeneracy of the $^{4}$T$_{1g}$ term resulting in an orbital singlet ground state. The effect of zero-field splitting is to produce two doublets, labelled by the $M_{S}$ component of spin, $S$. Neutron diffraction and ESR determined the $M_{S} = \pm 3/2$ state to be lowest in energy. \label{highspin_co}}
\end{figure}

For high-spin Co(II), the orbital contribution to the moment and the presence of the ligand field allows for strong magnetic anisotropy. The energy levels of a single Co(II) ion (3d$^{7}$, $^{4}$F) are split for a six-coordinated Co(II) ion in an octahedral environment into three terms, the lowest of which is the orbital-triplet state $^{4}$T$_{1g}$.\cite{ZFS, Co1} The low temperature structure of [Co(HF$_{2}$)(pyz)$_{2}$]SbF$_{6}$ suggests that the CoN$_{4}$F$_{2}$ octahedra are compressed along the Co$-$F bond, and this distortion lifts the degeneracy of the $^{4}$T$_{1g}$ term,\cite{Co1, Co2} producing an orbital-singlet ground state, $^{4}$A$_{2g}$  (Fig. \ref{highspin_co}). The leading term in the crystalline-electric field (CEF) operators for an axially compressed octahedral environment takes the form $\mathcal{H}_{\textsc{cef}}$ =$-DS_{z}^{2}$,\cite{CEF} and this zero-field splitting partially lifts the degeneracy of the four spin-states, resulting in two doublets separated by 2$D$. From the anisotropy of the ordered moments deduced from neutron diffraction, we unambiguously determine that the zero-field splitting in this polymeric magnet is Ising-like, resulting in the $M_{S}$ = $\pm$~3/2 ground-state doublet. At temperatures low compared to 2$D$, only the doublet ground state is thermally occupied and these two levels then dominate the magnetic properties of the material.

\subsubsection{Heat Capacity} 
 
\begin{figure}[b]
\includegraphics[width=\columnwidth]{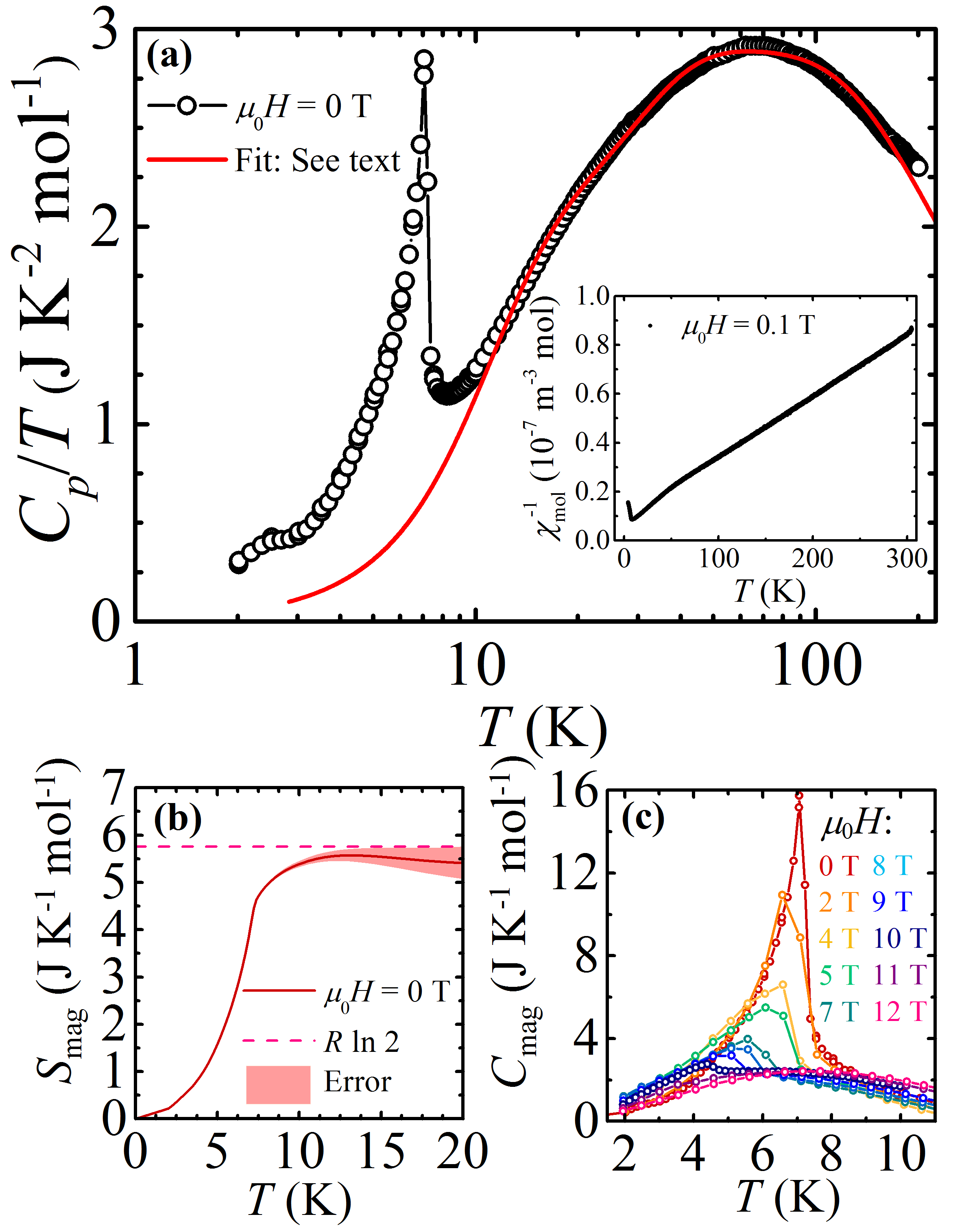}
\caption{\textbf{(a)} The ratio of heat capacity ($C_{p}$) to temperature ($T$) as a function of $T$ for $[$Co(HF$_{2}$)(pyz)$_{2}$$]$SbF$_{6}$, collected in zero applied field. The solid line shows the result of fitting the data to a model of the lattice contribution (described in text). Inset: Inverse susceptibility ($\chi_{\text{mol}}^{-1}$) vs. temperature for $[$Co(HF$_{2}$)(pyz)$_{2}$$]$SbF$_{6}$, recorded with $\mu_{0}H$ = 0.1~T. \textbf{(b)}~Entropy change vs. temperature deduced by integrating the $C_{\text{mag}}$ using Eq.~\ref{entropy_calc}. The shaded region indicates the error on this calculation due to uncertainties in the subtraction of the lattice fit.  \textbf{(c)} Magnetic contribution to the heat capacity vs. temperature at representative fields in the range 0 $\le \mu_{0}H \le$ 12 T. \label{CoSbF6_hc_fit}}
\end{figure}

The measured heat capacity ($C_{p}$) of $[$Co(HF$_{2}$)(pyz)$_{2}$$]$SbF$_{6}$ is plotted as a function of temperature in Fig. \ref{CoSbF6_hc_fit}(a). The data shows a sharp $\lambda$-peak at $T_{\textsc{n}}$ = 7.1 K indicating a magnetic transition to an ordered state, superimposed on a sloping background due to the phonon contribution.

In the case of the quasi-two dimensional system [Cu(HF$_{2}$)(pyz)$_{2}$]SbF$_{6}$, short-range correlations of spins within the planes that build up above $T_{\textsc{n}}$ give rise to a broad maximum the heat capacity and consequently a small peak in $C_{p}$ at the transition to long range order, which occurs at temperature less than $J_{\parallel}/k_{\textsc{b}}$.\cite{Cu_sat_moment} Simulations\cite{Pinaki} predict that as the magnetic exchange between layers is increased in these Heisenberg systems, such that the inter- and intra-plane interactions become more equal, the transition temperature increases towards $J_{\parallel}/k_{\textsc{b}}$ and the appearance of the heat capacity approaches a single large $\lambda$-peak at $T_{\textsc{n}}$. However, to explain the form of the heat capacity for [Co(HF$_{2}$)(pyz)$_{2}$]SbF$_{6}$ we must also consider the consequences of the result from neutron diffraction, which indicates the Co(II) moments have Ising character. For both square-lattice 2D Ising\cite{IsingHC2D} and simple-cubic 3D Ising\cite{IsingHC3D} systems, the heat capacity exhibits large single peak at a finite transition temperature, whilst the broad maximum indicative of correlations building up above $T_{\textsc{n}}$ is only recovered for well isolated 1D chains.\cite{IsingHC2D}  We therefore attribute the appearance of the observed $\lambda$-peak in the Co(II) material primarily to the Ising nature of the Co(II) moments and secondarily conclude that the data are consistent with 2D or close to 3D spatial exchange anisotropy.

A quantitative estimate of the entropy change associated with the transition was found by subtracting the contribution to $C_{p}$ from phonons. This was performed by first fitting data from 11 $\le T \le$ 200 K to an expression for the lattice contribution ($C_{\text{latt}}$) given by a sum of one Debye mode and two Einstein modes\cite{Cu_sat_moment} [We note that this fit absorbs the broad Schottky anomaly feature expected in the heat capacity from thermal depopulation of the crystalline-electric field split energy levels (Fig. \ref{highspin_co})]. The resulting fitted parameters are summarised in Table \ref{hc_fit}, and compared to those of $[$Cu(HF$_{2}$)(pyz)$_{2}$$]$SbF$_{6}$. The energy scale of each mode is similar for both lattices, most likely due to the shared structure giving rise to similar phonon contributions to $C_{p}$.

The entropy change ($\Delta S_{\text{mag}}$) at the transition temperature is deduced by numerically integrating

\begin{equation}
\Delta S_{\text{mag}} =   \int_{0}^{\infty}\frac{C_{\text{mag}}}{T} \text{ d}T,
\label{entropy_calc}
\end{equation}

\begin{table}[t]
 \caption{Fitted lattice contribution to the heat capacity for $[$Co(HF$_{2}$)(pyz)$_{2}$$]$SbF$_{6}$ compared to the published results for $[$Cu(HF$_{2}$)(pyz)$_{2}$$]$SbF$_{6}$. \cite{Cu_sat_moment} $A_{i}$ = amplitude of mode $i$, $\theta_{i}$ = characteristic Temperature of mode $i$ ($i$ = D, Debye; E, Einstein). \label{hc_fit}}
 \begin{ruledtabular}
 \begin{tabular}{l r r}
Parameter & $[$Co(HF$_{2}$)(pyz)$_{2}$$]$SbF$_{6}$ & $[$Cu(HF$_{2}$)(pyz)$_{2}$$]$SbF$_{6}$ \\
(units) & & \\
\hline
$A_{\text{D}}$ (J/K.mol) & 82(1) & 76(1) \\
$\theta_{\text{D}}$ (K) & 80(1) & 94(8) \\
$A_{\text{E$_{1}$}}$ (J/K.mol) & 174(3) & 134(4) \\
$\theta_{\text{E$_{1}$}}$ (K) & 177(2) & 208(4) \\
$A_{\text{E$_{2}$}}$ (J/K.mol) & 287(3) & 191(4) \\
$\theta_{\text{E$_{2}$}}$ (K) & 448(6) & 500(3) \\
 \end{tabular}
 \end{ruledtabular}
 \end{table}

\noindent
where $C_{\text{mag}} = C_{p} - C_{\text{latt}}$. Taking $C_{p}$($T$ = 0 K) = 0, the resultant entropy change associated with the $\lambda$-peak, (from 0 $\le T \le$ 20 K) is $\Delta S_{\text{mag}}$ = 5.4(3) J~K$^{-1}$~mol$^{-1}$, [Fig. \ref{CoSbF6_hc_fit}(b)]. This is consistent with the magnetic ordering of a two-state system, $\Delta S_{\text{mag}}$ = $R_{ }\ln 2$ = 5.8 J K$^{-1}$ mol$^{-1}$, which indicates that the Co(II) ions act as a two-level system below 20 K. This is in agreement with the conclusion from neutron powder diffraction at 4~K that the Co(II) ions exhibit a ground-state doublet with Ising character. In addition, the result that the anisotropy persists above $T_{\textsc{n}}$ further supports the suggestion that the anisotropy in this material relates to the single-ion properties of each Co(II) moments. The inverse susceptibility, 1/$\chi_{\text{mol}}$, departs from its linear high temperature behaviour below approximately 50 K [Fig.~\ref{CoSbF6_hc_fit}(a)]. If the exchange energy is small, as is the case here, this departure occurs in the paramagnetic phase at an energy scale comparable to typical values of $D$ in octahedral Co(II) complexes.\cite{Co2} We therefore attribute this feature to the thermal depopulation of CEF split levels, resulting in the observed entropy change at $T_{\textsc{n}}$.

The field dependence of the heat capacity was investigated in applied fields up to $\mu_{0}H =$ 12 T [Fig. \ref{CoSbF6_hc_fit}(c)]. The $\lambda$-peak is suppressed in height, and moves to lower temperature as the applied field is increased, as expected for a transition to an antiferromagnetically ordered state.

We note that there is a small, as yet unexplained, anomaly in the zero-field heat capacity at $\approx$ 2.5 K [Fig. \ref{CoSbF6_hc_fit}(a)], which becomes very weak in an applied field of $\mu_{0}H$ = 1 T, and is not visible above the background for $\mu_{0}H \ge$ 2 T. Since the neutron diffraction determined the ordered magnetic ground state at 4 K, the anomaly does not change the conclusion that the large peak at $T_{\textsc{n}}$ is associated with a transition to a commensurate long-range ordered phase. The neutron diffraction did detect a small impurity phase (Fig. \ref{CoSbF6_neutrons}), and this may be associated with the small anomaly in the heat capacity.

\subsubsection{ESR}

\begin{figure}[t]
\includegraphics[width=\columnwidth]{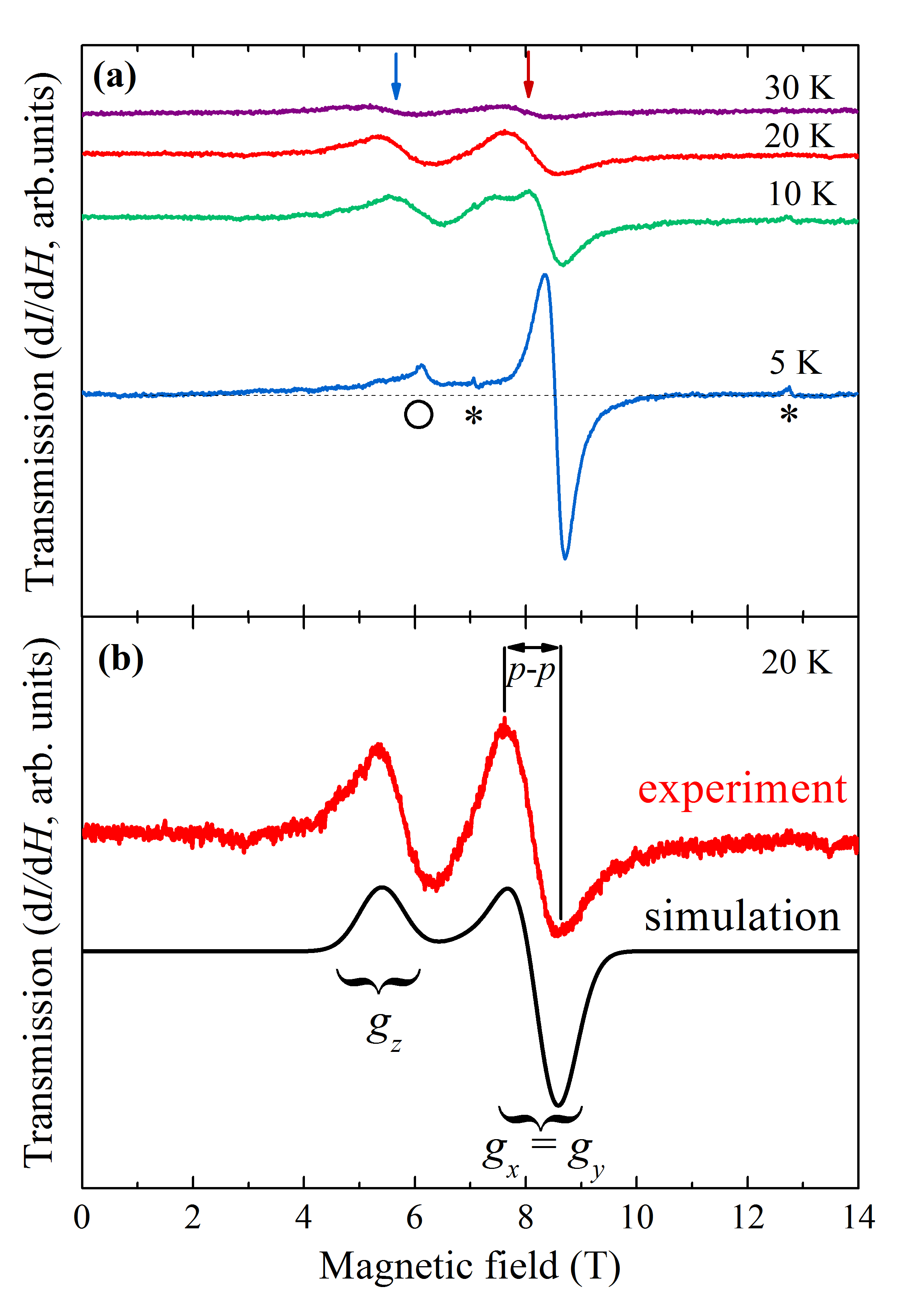}
\caption{\textbf{(a)} Transmission ESR spectra (in derivative mode) vs. applied field at 416~GHz. At 30~K, two transitions are seen due to single Co(II) ion resonances with anisotropic $g$-tensor. Below $T_{\textsc{n}}$ = 7.1 K a large AFM resonance is observed, and the frequency dependence of this mode is shown in Fig. \ref{ESR3}. The asterisk marks paramagnetic impurities and $\bigcirc$ is attributed to a critical resonance mode at the spin-flop field. \textbf{(b)} The experimental (red) and simulated (black) 20~K ESR spectra at 416 GHz. The simulation was performed with an effective spin-1/2 paramagnetic model with an anisotropic $g$-factor: $g_{x}$ = $g_{y}$ = 3.6(4), $g_{z}$ = 5.5(7); and an isotropic line width (defined as the FWHM) of 800 mT.\label{ESR2}}
\end{figure}

ESR spectra measured at fixed temperatures in the range 5 $\le T \le$ 30 K and in fields $\mu_{0}H \le$ 14 T, are shown in Fig. \ref{ESR2}(a). From 20 K to 30 K (in the paramagnetic phase) two $T$-independent transitions are found, which we associate with single spin excitations of Co(II) ions.

At temperatures and fields low compared to the energy scale of 2$D$ (Fig. \ref{highspin_co}), the Co(II) ions can treated as a two-level system, with an anisotropic $g$-tensor. The structure suggests two principle $g$-factor axes, taken to be parallel and perpendicular to the [Co(pyz)$_{2}$]$^{2+}$ sheets (the $xy$-plane and $z$-axis respectively). Magnetic exchange in Co systems can be treated within this model using the effective spin-half Hamiltonian,\cite{spin_half1, spin_half2} given by

\begin{multline}
\hat{\mathcal{H}} = J\sum_{\langle i,j \rangle}\left[\hat{S}_{i}^{z}\hat{S}_{j}^{z}+ a \left( \hat{S}_{i}^{x}\hat{S}_{j}^{x}+\hat{S}_{i}^{y}\hat{S}_{j}^{y} \right)\right]\\ + \sum_{i}\mu _{\textsc{b}}\mu _{0}\textit{\textbf{H}}\cdot \textit{\textbf{g}}\cdot \hat{\mathbf{S}}_{i},
\label{ham}
\end{multline}

\noindent
where $\textit{\textbf{g}}$ = diag($g_{\text{xy}}$, $g_{xy}$, $g_{z}$), $\hat{\mathbf{S}}$ = ($\hat{S}_{i}^{x}$, $\hat{S}_{i}^{y}$, $\hat{S}_{i}^{z}$) are the spin-half operators and ${\langle i,j \rangle}$ indicates a sum over unique pairs of Co(II) ions. Here, $J$ is the strength of the magnetic exchange to $N$ nearest neighbours and the spatial exchange anisotropy for this material will be discussed in Section IV. The size of the true Co(II) moment is absorbed into the $g$-factor and $J$, whilst the direction of the moments due to the easy-axis anisotropy is represented by the spin-exchange anisotropy ($0\le a \le 1$) and the anisotropic $g$-factor.\cite{spin_half1} 

For temperatures above $T_{\textsc{n}}$ but below the energy scale of 2$D$ such that the effective spin half model applies, the Zeeman term in Eq.~\ref{ham} can be used to simulate the expected ESR spectra and this is compared to data in Fig. \ref{ESR2}(b). This model predicts two features: (i) a small peak for the $g_{z}$ transition; and (ii) a larger derivative shape transition for the $g_{xy}$ transition. Given the stronger signal for the higher field transition in the measured data, we attribute this to the $xy$-transition. 

The measured fields at which each transition occurs indicates $g_{z}$ = 5.5(7) and $g_{xy}$ = 3.6(4), where the error derives from the finite width of the broad ESR modes, which leads to an uncertainty in the critical fields used to determine the $g$-factors. This implies that size of the moments along the $c$-axis is $g_{z}\mu_{\textsc{b}}S_{\text{eff}}$ =  2.8(4)$\mu_{\textsc{b}}$ (where $S_{\text{eff}}$ = 1/2 is the effective spin quantum number), showing that the moment per ion in the paramagnetic phase is consistent with the ordered moment deduced from neutron diffraction, $\mu_{\text{Co}}$ = 3.02(6)$\mu_{\textsc{b}}$. In addition, the relative sizes of the $g$-factors determined from ESR is in agreement with the presence of strong local easy-axis anisotropy of the each Co(II) moment. This supports the conclusion from heat capacity measurements that the anisotropy persists above $T_{\textsc{n}}$, indicating that it originates from single-ion properties of the Co(II) moments due to their interaction with the crystalline electric field.

\begin{figure}[t]
\includegraphics[width=\columnwidth]{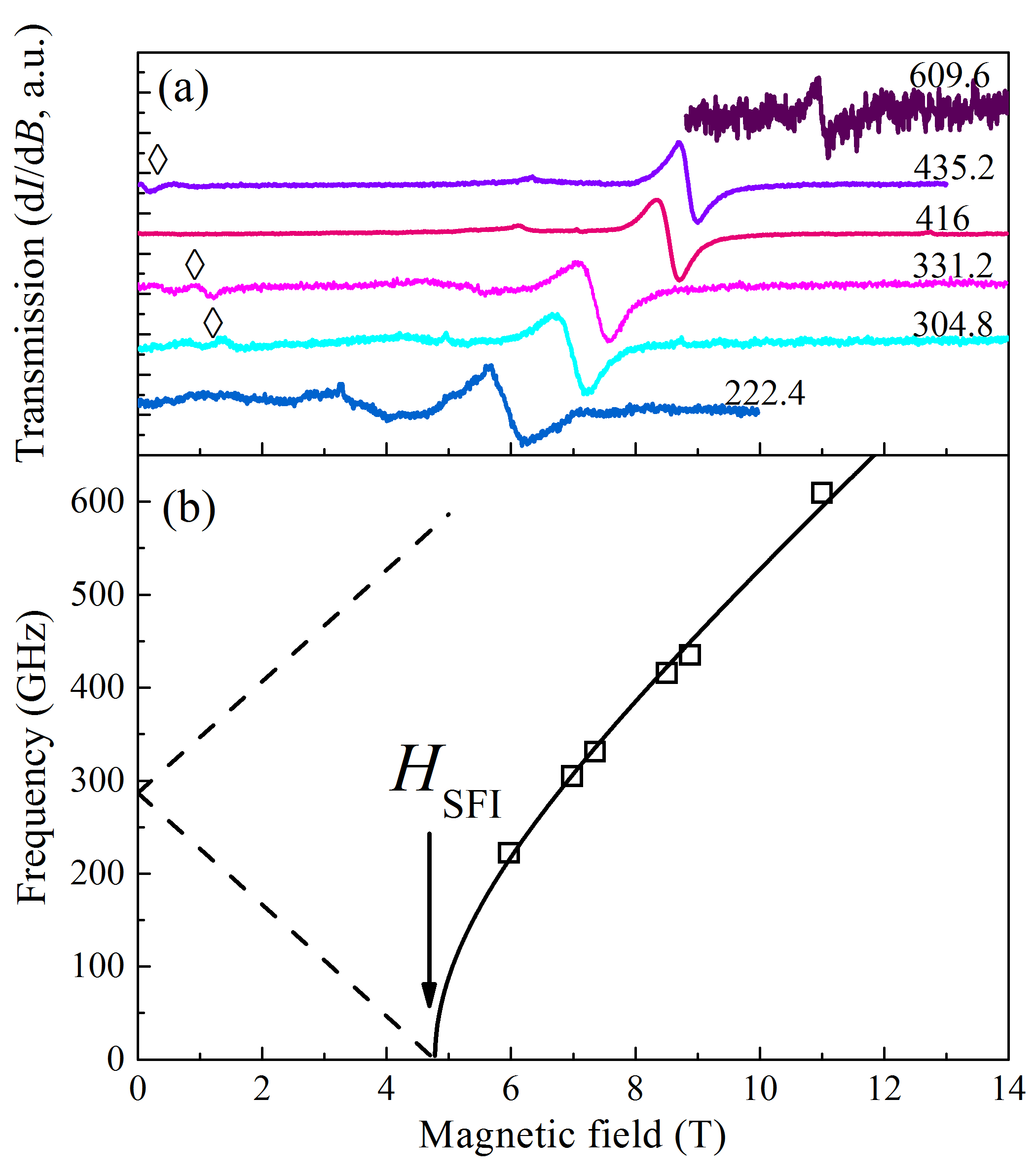}
\caption{\textbf{(a)} Transmission ESR spectra for [Co(HF$_{2}$)(pyz)$_{2}$]SbF$_{6}$ as a function of applied field, for frequencies in the range 222.4 $\le \nu \le$ 609.6~MHz. \textbf{(b)}~Frequency dependence of ESR spectra recorded at 5~K, showing the experimentally measured antiferromagnetic resonances ($\Box$) compared to those expected from simulations (lines). The solid and dashed lines (simulated with a powder-average $g$ = 4.21 and $H_{\text{sfI}}$ = 4.9~T) correspond to the ESR transitions for applied fields above and below $H_{\text{sfI}}$. Transitions corresponding to the dashed lines ($H \le H_{\text{sfI}}$) are weak, however features indicated with diamonds in panel (a) follow a linear frequency-field dependence consistent with $H_{\text{sfI}}$ = 4.9~T and  $g_{z} \approx$ 6.3. This larger $g$-factor is close to the full moment measured by neutron diffraction in zero-field, which suggests these resonances correspond to the low-field state with the cobalt moments aligned to the $z$-axis.\label{ESR3}}
\end{figure}

The transition width (labelled as $p$-$p$ in Fig. \ref{ESR2}) is 100 times wider than that observed in [Cu(HF$_{2}$)(pyz)$_{2}$]SbF$_{6}$. In Heisenberg exchange coupled systems, narrow transitions are expected,\cite{EPR1} so the broad observed resonances suggest the presence of non-Heisenberg interactions.\cite{EPR2}

On cooling the sample to below $T_{\textsc{n}}$, a large antiferromagnetic resonance mode is seen, exhibited by the derivative shape in the 5 K data in Fig. \ref{ESR2}(a). The frequency dependence of this mode was recorded for 6 frequencies ($\nu$) in the range 200 $\le \nu \le$ 600 GHz, and was found to move to higher fields with increased $\nu$ (Fig. \ref{ESR3}). An easy-axis AFM in the ordered phase is expected to undergo a field-induced spin-flop transition at a field $H_{\text{sf}}$. The field dependence of the observed antiferromagnetic resonance is fitted to the equation\cite{EPR4}

\begin{equation}
\frac{h\nu}{g\mu_{\text{B}}} = \mu_{\text{0}}\sqrt{H^{2}-H_{\text{sfI}}^{2}}.
\label{H_sf}
\end{equation}
\noindent
for $H \ge H_{\text{sfI}}$, where $H_{\text{sfI}}$ is a lower bound of the spin-flop field which is realised in the case of ideal Ising spins. The resultant fitted parameters are $H_{\text{sfI}}$ = 4.9 T and $g$~=~4.21, which is in agreement with the powder average $g$-factor of the effective spin-half model at 20 K.

Below $T_{\textsc{n}}$, a frequency independent critical field resonance mode is expected as the field is swept through the spin-flop field.\cite{EPR4} At 5 K, we observe an ESR mode that begins to emerge at an applied field of $\mu_{0}H \approx$ 5 T, leading to a small peak at 6.1(1) T [labelled $\bigcirc$ in Figure \ref{ESR2}(a)]. This mode was found to be $\nu$-independent so we attribute this to the critical resonance mode at the spin-flop field. In addition, the strongest ESR resonance vanishes below 6 T, which is expected to occur when the applied field is less than the spin-flop field\cite{EPR5} and is further consistent with a $\mu_{0}H_{\text{sf}} \lesssim$ 6 T.

The critical resonance was only observed over a narrow frequency range, so to further explore the behaviour of the spins for applied fields in vicinity of this mode, magnetization measurements were recorded at various temperatures through this region.

\subsubsection{Magnetization}

\begin{figure}[b]
\includegraphics[width=\columnwidth]{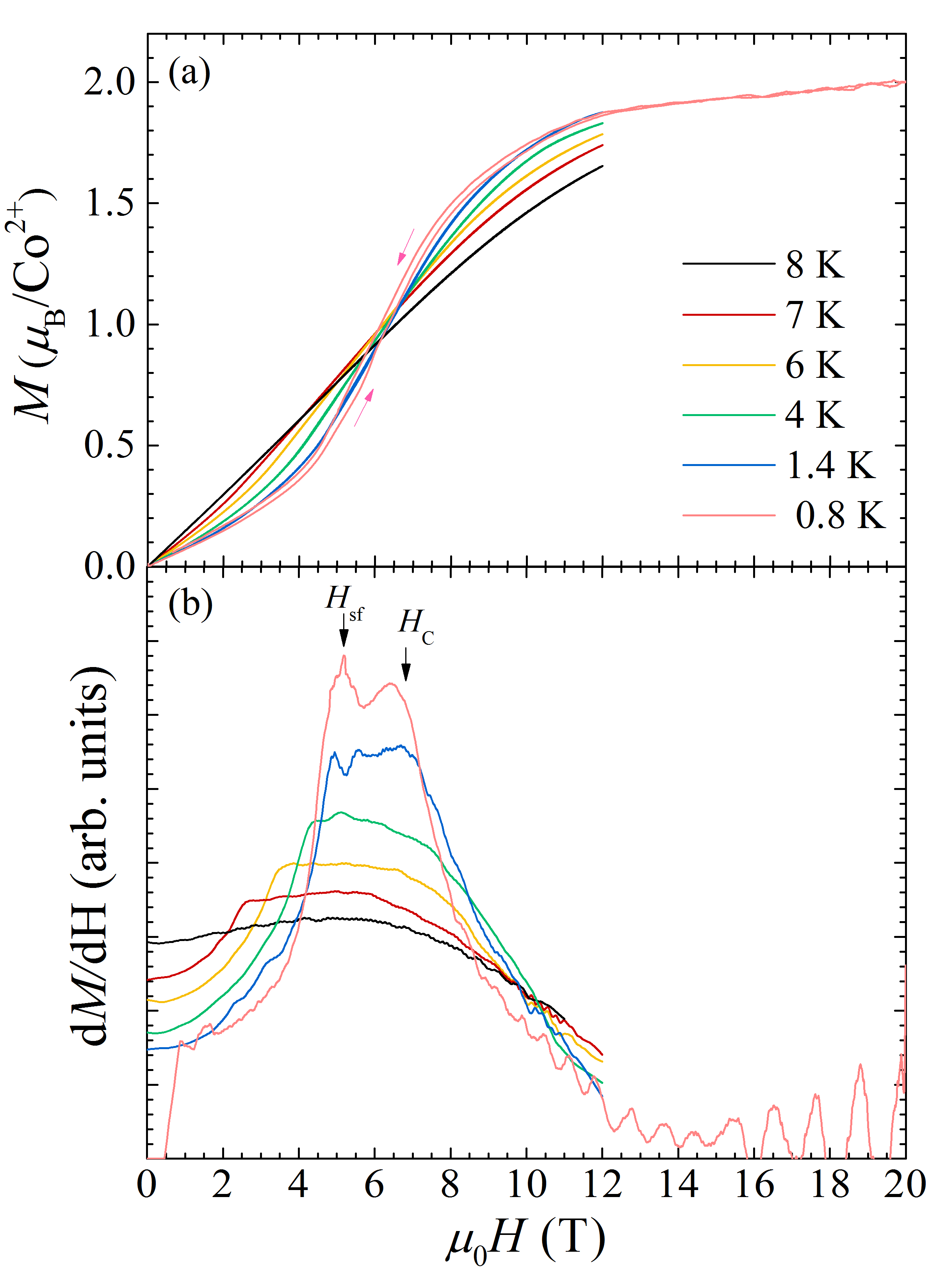}
\caption{\textbf{(a)} Magnetization vs. applied field for a polycrystalline sample of $[$Co(HF$_{2}$)(pyz)$_{2}$]SbF$_{6}$. As the temperature is lowered below $T_{\textsc{n}}$ = 7.1 K, a low-field kink in the magnetization develops, which tends to move to higher fields on cooling. Data at 0.8 K were recorded with pulsed fields, and the arrows indicate the direction of the field sweep. \textbf{(b)} d$M$/d$H$ vs. applied field. Data collected at 0.8 K shows the derivative of the down-sweep measured from pulsed-field magnetization. On cooling, we see the development of two features in the magnetization, labelled at $H_{\text{sf}}$ and $H_{c}$.\label{CoSbF6_m_vs_B}}
\end{figure}

A quasi-static measurement of magnetization for fixed temperatures in the range 1.4 $\le T \le$ 8 K was performed with a VSM, and the resulting curves are shown in Fig.~\ref{CoSbF6_m_vs_B}. Above $T_{\textsc{n}}$, the magnetization rises smoothly, but below $T_{\textsc{n}}$ a kink develops at low fields resembling a spin-flop transition as implied from ESR (see above). This feature was found to move to higher fields as the sample was cooled. In addition, for temperatures below $T_{\textsc{n}}$, magnetization sweeps  appear to cross each other at $\mu_{0}H$ = 6.5(2) T.

Pulsed-field magnetization data for $T \ge 0.8$ K and $\mu_{0}H \le$ 20~T were also recorded, and a representative curve at 0.8 K is added to Fig. \ref{CoSbF6_m_vs_B}(a). Using the VSM data to normalise the pulsed-field magnetization, we find that the moment of the sample parallel to the field at 20~T reaches $\approx$~2~$\mu_{\textsc{b}}$. This bulk measurement of the moment from a powdered sample is much less than the ordered moment as measured from the local probe of neutron diffraction, which is in agreement with the proposal for the presence of strong single-ion anisotropy in this sample. Furthermore, the moment at 20~T is consistent with the powder average value of $g\mu_{\textsc{b}}S_{\text{eff}}$ = 2.2(4)$\mu_{\textsc{b}}$ as implied by ESR, and suggests that effective spin half model applies up to at least 20~T at low temperatures in this sample.

Given this, we use the Hamiltonian in Eq.~\ref{ham} to model the system through the spin-flop phase. In order to apply this Hamiltonian to our system, we represent the moment of each Co(II) with a vector. This approximation is justified in the case of [Co(HF$_{2}$)(pyz)$_{2}$]SbF$_{6}$ by considering two differences of the Co(II) system from the Cu(II) system: (i) the Co(II) ions have a higher spin quantum number ($S$ = 3/2); and (ii) the same moment size per ion is measured in the paramagnetic and ordered phases, from ESR and neutron diffraction respectively.

In zero field, Eq.~\ref{ham} (with $a < 1$) gives AFM ordered moments parallel to $z$. For a field applied along $z$, initially the spin-exchange anisotropy leads to no response in the magnetization. As the field increases, a spin-flop transition occurs  at $H$ = $H_{\text{sf}}$, at which point the component of the Co(II) moments within the [Co(pyz)$_{2}$]$^{2+}$ $xy$-planes are antiferromagnetically ordered whilst the spin-component along $z$ becomes ferromagnetically aligned, causing d$M$/d$H$ to increase sharply. The moments then continue to cant towards the applied field as it is increases further, until saturation occurs at $H = H_{\text{c}}$, with the collinear moments fully aligned to the z direction. For a field applied perpendicular to $z$, the moments slowly cant towards the direction of the field $H$, and will approach saturation at $H > H_{\text{c}}$. In a powdered sample, d$M$/d$H$ will decrease for an applied field above $H_{\text{c}}$ since magnetic response is only from the portion of the sample for which the field is not aligned with the local easy-axis of the Co(II) ions.

The measured powder differential susceptibility (d$M$/d$H$) [Fig. \ref{CoSbF6_m_vs_B}(b)] shows the fine structure of the magnetization around the spin-flop transition. Below $T_{\textsc{n}}$, d$M$/d$H$ increases to a sharp transition at which the differential susceptibility approximately plateaus. As the field is increased further, a second transition is reached at which point d$M$/d$H$ decreases. We attribute the two fields at the edges of the plateau to $H_{\text{sf}}$ and $H_{\text{c}}$. At 0.8~K, $\mu_{0}H_{\text{sf}}$~=~5.1(2)~T and $\mu_{0}H_{\text{c}}$~=~6.8(2)~T.

A quantitative estimate of $J$ and $a$ can be made by comparing these two measured values of the critical fields to those calculated by applying Eq.~\ref{ham} to the model where each cobalt moment is represented as a vector. Within this approximation, we find the critical fields are given by
\begin{equation}
g_{z}\mu_{\textsc{b}}\mu_{0}H_{\text{sf}} = NJ(1-a^{2})^{\frac{1}{2}}S_{\text{eff}}
\label{mag_1}
\end{equation}
and
\begin{equation}
g_{z}\mu_{\textsc{b}}\mu_{0}H_{\text{c}} = NJ(1+a)S_{\text{eff}},
\label{mag_2}
\end{equation}
\noindent
where $S_{\text{eff}}$ is the effective spin, $S_{\text{eff}}$ = 1/2. Given the conclusion from heat capacity measurements that the spatial exchange anisotropy may be 2D or 3D, we begin by assuming that the inter$-$ and intra-plane exchange interactions in this material are of the same order ($J_{\parallel}$ = $J_{\perp}$ = $J$), and take $N$~=~6. Using the $g$-factor deduced from ESR, we find from Eq.~\ref{mag_1} and \ref{mag_2} that the magnetization data can be represented with $a$ = 0.28(6) and an effective exchange parameter $NJ/k_{\textsc{b}}$ = 39(9) K, yielding $J/k_{\textsc{b}}$~=~7(1)~K. We examine this result in two limits. For Heisenberg $S$ = 1/2 moments ($a$ = 1) on a simple cubic lattice, $k_{\textsc{b}}T_{\textsc{n}} \approx 0.9J$, whilst within the effective $S_{\text{eff}}$~=~1/2 Ising model (with $a$ = 0) for moments on a simple cubic lattice, $k_{\textsc{b}}T_{\textsc{n}} \approx 1.1J$.\cite{Carlin, note} Using the transition temperature determined from heat capacity and the value of $J$ derived above, we find that for [Co(HF$_{2}$)(pyz)$_{2}$]SbF$_{6}$ $k_{\textsc{b}}T_{\textsc{n}}$~=~1.1(2)$J$, which indicates that the proposed values of $a$~=~0.28(6) and $J/k_{\textsc{b}}$~=~7(1) K are consistent with the simple cubic model.

We note that the strength of the magnetic interactions in related polymers can vary. The tetragonal compound CoCl$_{2}$(pyz)$_{2}$ also consists of square [Co(pyz)$_{2}$]$^{2+}$ planes,\cite{CoCl2} and has a measured transition temperature of 0.86(2)~K.\cite{CoCl2pyz2} This system is also proposed to have 2D or 3D exchange interactions and, but in contrast to our material takes on $XY$ (easy-plane) anisotropy of the Co moments.\cite{CoCl2pyz2} Clearly, the local Co(II) environment is an important factor in determining the magnetic properties of these systems.

\section{Discussion}

The ordered moment in [Cu(HF$_{2}$)(pyz)$_{2}$]SbF$_{6}$ at 2 K,  $\mu_{\text{Cu}}$ = 0.6(1)$\mu_{\textsc{b}}$, is much less than the moment expected,\cite{Cu_sat_moment} $g\mu_{\textsc{b}}S$ = 1.07(1)$\mu_{\textsc{b}}$, for $S$ = 1/2. This suggests the presence of strong quantum fluctuations acting on the ground state, and the degree to which the moment is reduced is in agreement with that predicted by calculations of single 2D layers of AFM $S$ = 1/2 Heisenberg moments on a square-lattice.\cite{SLAFM} This reduction in moment is also comparable to that observed in other low-dimensional antiferromagnetic Cu(II) systems despite the relatively high value of $k_{\textsc{b}}T_{\textsc{n}}/J_{\parallel}$, which parameterises the dimensionality of the system, and a comparison to related systems is given in Table \ref{reduced_moment}.

\begin{table}[b]
 \caption{Ordered moments compared to $k_{\textsc{b}}T_{\textsc{n}}/J_{\parallel}$ for three related quasi-2D systems, and the last example is of a quasi-1D chain. A smaller value of the ratio $k_{\textsc{b}}T_{\textsc{n}}/J_{\parallel}$ indicates a lower dimensionality of the  magnetic system. The column labelled method refers to the experimental technique from which the moment size estimates were deduced: PND = powder neutron diffraction, SCND = single crystal neutron diffraction and $\mu^{+}$SR = muon-spin rotation. \label{reduced_moment}}
 \begin{ruledtabular}
 \begin{tabular}{l l l l l l}
Compound & $\mu_{\text{Cu}}$ ($\mu_{\textsc{b}}$) & ref. & $k_{\textsc{b}}T_{\textsc{n}}/J_{\parallel}$ & ref. &Method\\
\hline
$[$Cu(HF$_{2}$)(pyz)$_{2}$]SbF$_{6}$ & 0.6(1) & & 0.32 & \cite{Cu_sat_moment} & PND \\
CuF$_{2}$(H$_{2}$O)$_{2}$(pyz) & 0.60(3) & \cite{Cu_neutrons} & 0.21 &\cite{CuF2} & SCND \\
Cu(pyz)$_{2}$(ClO$_{4}$)$_{2}$ & 0.47(5)  & \cite{CuClO4} & 0.23 & \cite{NJP} & SCND \\
Cu(pyz)(NO$_{3}$)$_{2}$ & 0.05 & \cite{TL_calc} & 0.01 & \cite{TL_calc} & $\mu^{+}$SR\\
 \end{tabular}
 \end{ruledtabular}
 \end{table}

A second measure of the moment of each Cu(II) ion can be estimated from the saturated moment in high fields. Using the single-crystal magnetization data taken at 2~K (Fig. \ref{CuSbF6_mvsB}) to calibrate the published pulsed-field data,\cite{Cu_sat_moment} we find that the saturated moment at low temperatures is $\mu_{\text{Cu}}$ = 1.1(1)$\mu_{\textsc{b}}$. This is equal to the full moment, and shows that whilst quantum fluctuations are present in the ground state, all fluctuations are suppressed in high fields. This property of the quantum fluctuations is also found directly in the related system Cu(pyz)$_{2}$(ClO$_{4}$)$_{2}$.\cite{CuClO4}

The renormalization of the magnetic moment in the ordered phase can be visualised with spin-wave perturbation theory.\cite{SLAFM} In this model, a spin-flip excitation can propagate though the system and interacts with the ground state of the magnet. This causes further excitations to be created which will annihilate at a later time, leaving the original spin flip excitation behind. The sum of all possible interaction processes of this type, which can occur in the propagation of an excitation between two points, are responsible for the observed renormalization of the ordered moment of the system.

Additionally, for [Cu(HF$_{2}$)(pyz)$_{2}$]SbF$_{6}$, we have found no evidence of a kink in the single crystal magnetization for any direction of applied magnetic field that would indicate a presence of non-Heisenberg interactions,\cite{Xiao} which would contribute to the finite ordering temperature.\cite{Cuccoli} The absence of a kink is an indication that the ordering temperature will be predominantly determined by the size of the interlayer coupling. The fact the neutron data showed that the moments in adjacent layers are antiferromagnetically aligned provides experimental evidence that the magnetic exchange along this direction, $J_{\perp}$, is antiferromagnetic in nature. If the finite $T_{\textsc{n}}$ is entirely due to interplane exchange, we find $J_{\perp}/k_{\textsc{b}}$ =~0.12~K.

Given this result, we can anticipate the largest size of the correlation length that can be sustained in the Q2D phase of [Cu(HF$_{2}$)(pyz)$_{2}$]SbF$_{6}$ before a transition to long range order occurs. Using the published\cite{Cu_sat_moment} $T_{\textsc{n}}$~=~4.3~K, and the measured reduction in ordered moment from the deuterated sample, Eq.~\ref{TN} predicts the material will order once the correlation length (in units of the lattice constant) reaches $\xi \approx 10 a$.

The magnetic properties of the Co system show contrasts to the Cu analogue due to four four key differences between these materials: (i) the full spin quantum number per ion is increased from $S$ = 1/2 in the Cu(II) material to $S$ = 3/2 in the Co(II) system; this in turn allows for two further distinctions: (ii) both the $d_{x^{2}-y^{2}}$ and $d_{z^{2}}$  orbitals contain unpaired electrons for Co(II) in the high-spin state, allowing for significant spin-density delocalization along both the pyrazine \textit{and} linear HF$_{2}^{-}$ ions and (iii) the Co(II) moments have Ising-like anisotropy. Lastly, there is a structural comparison (iv) the plane spacing in the Co sample is smaller compared to that in the Cu sample.

Differences (ii) and (iv) are likely to promote stronger magnetic coupling between ions in adjacent layers in the Co(II) system. Since the F$\cdots$H$\cdots$F ligand is known to be an effective mediator of magnetic exchange in the isostructural [Ni(HF$_{2}$)(pyz)$_{2}$]SbF$_{6}$,\cite{NiSbF6} we therefore expect that the ratio of the inter- to intra-plane exchange, $J_{\perp}/J_{\parallel}$ to be increased compared to [Cu(HF$_{2}$)(pyz)$_{2}$]SbF$_{6}$. The heat capacity is consistent with this ratio being increased until the strength of the exchange interactions become isotropic, since a single sharp peak is expected for 2D and 3D Ising systems. If the ratio were to be much greater than 1, the system would tend towards a new limit of a 1D chain of strongly coupled ions along the $c$-axis. For [Co(HF$_{2}$)(pyz)$_{2}$]SbF$_{6}$, the fact that no broad maximum was observed in the magnetic heat capacity suggests a 1D model of well isolated chains would not adequately describe the data.

The ordered magnetic moment of the Co(II) ions, $\mu_{\text{Co}}$~=~3.02(6)$\mu_{\textsc{b}}$, is comparable to the full moment measured from ESR for a temperature below the energy scale of the single ion anisotropy but above $T_{\textsc{n}}$, thus the effect of quantum fluctuations in the ordered phase is not observed in the neutron powder diffraction data at 4 K. This result is in contrast to the Cu(II) system and can be accounted for in terms of the comparisons outlined above.

Difference (i) is expected to have a small influence on the renormalized moment size. Linear spin-wave theory predicts that for Heisenberg $S$~=~1/2 moments on a square lattice the ordered moment is renormalised to 60$\%$ of its full value, whereas for Heisenberg $S$ = 3/2 moments on a square lattice, the moment is still renormalised to 87$\%$ of the full moment.\cite{Anderson} A stronger effect derives from comparison (iii) that the Co moments have Ising-character. A pure 2D Ising system orders at a finite temperature with the full moment\cite{2DIsing1} and, experimentally, spin-waves are not observed in 2D Ising systems,\cite{2DIsing2} removing the mechanism by which the moment is reduced in the Heisenberg model. Furthermore, 1D Ising systems are also expected to exhibit the full moment at $T$ = 0, and only show a reduced moment when the system is perturbed, for instance by the application of a transverse field.\cite{sachdev2}

In real quasi-1D systems consisting of high-spin Co(II), a reduction in the ordered moment is observed (although the reduction is less than in $S$ = 1/2 Heisenberg chains),\cite{1DIsing1, 1DIsing2} which can be attributed to the presence anisotropic spin-exchange interactions that in turn lead to a departure from the pure Ising limit. Magnetization data suggest such interactions are also present in [Co(HF$_{2}$)(pyz)$_{2}$]SbF$_{6}$, so the fact that the full moment is observed in the ordered phase is in keeping with the analysis of the spatial exchange anisotropy from the heat capacity data above. Therefore while differences (ii) and (iv) could lead to an increase in the exchange anisotropy $J_{\perp}/J_{\parallel}$ as compared to the highly 2D case of the Cu(II) system, it is not so large so as to move the system towards the limit of isolated 1D chains.

\begin{figure}[t]
\includegraphics[width=\columnwidth]{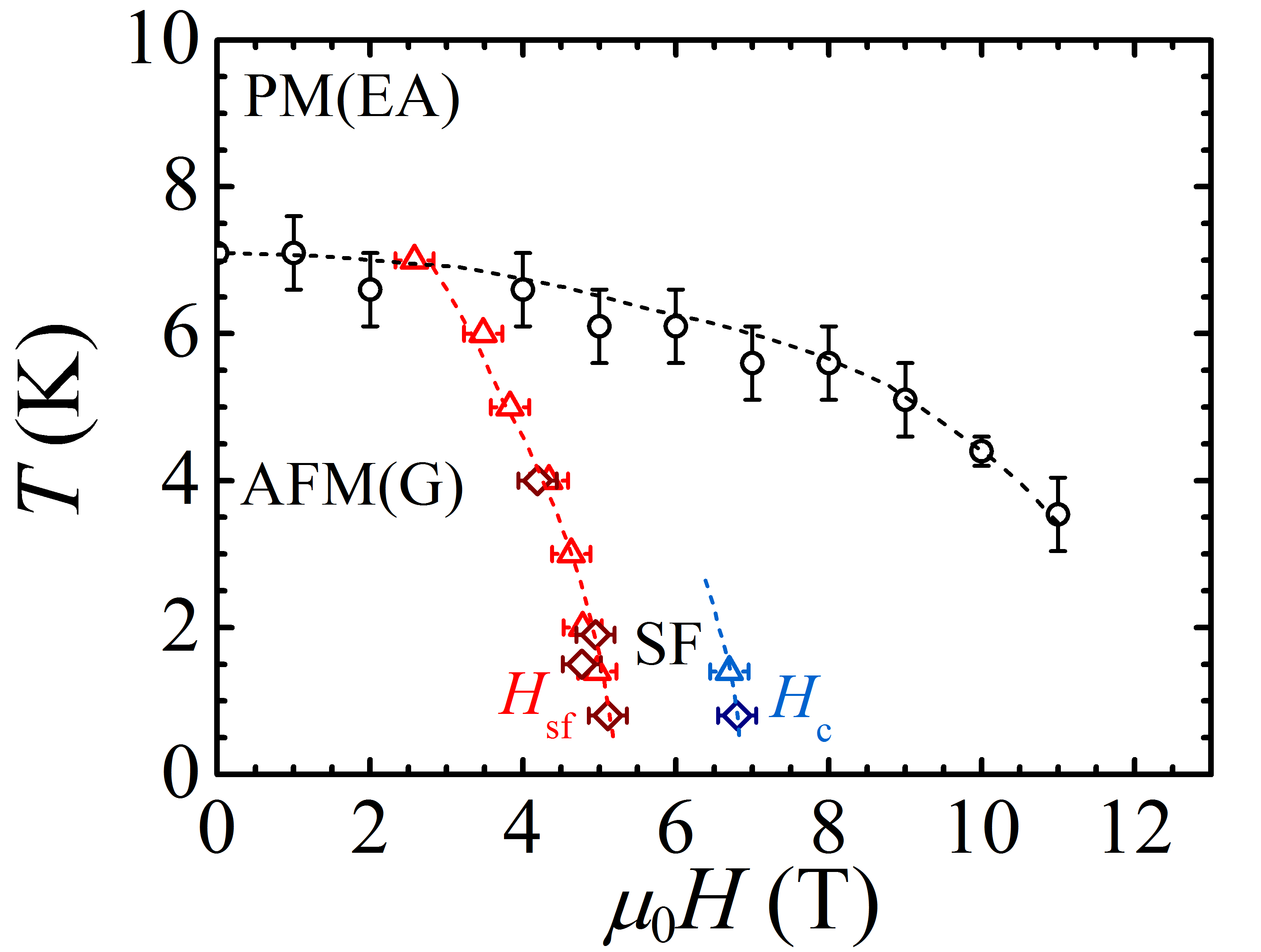}
\caption{Temperature-field phase diagram for [Co(HF$_{2}$(pyz)$_{2}$]SbF$_{6}$ showing the transition to long range order as measured by heat capacity (circles). The temperature dependence of the spin-flop ($H_{\text{sf}}$) and critical field ($H_{\text{c}}$) are also shown, as measured from DC magnetization (triangles) and pulsed-field magnetization measured on decreasing field (diamonds). AFM(G) = G-type antiferromagnetic order (deduced from zero-field neutron diffraction at 4 K); PM(EA) = Paramagnetic phase with easy-axis anisotropy; and SF = spin-flop phase. The dotted lines are a guide to the eye only. \label{CoSbF6_phase_diagram}}
\end{figure}

A summary of the magnetic phases of $[$Co(HF$_{2}$)(pyz)$_{2}$]SbF$_{6}$ are mapped out in a low-temperature vs. field phase diagram (Fig. \ref{CoSbF6_phase_diagram}). Above the ordering temperature, the paramagnetic ions exhibit single-ion easy-axis anisotropy, leading to effective spin-half behaviour to at least 30 K. On cooling, the  Co(II) moments enter a G-type antiferromagnetic ordered state, indicating that the magnetic exchange mediated by pyrazine and the HF$_{2}^{-}$ ligands are both antiferromagnetic in nature. On the subsequent application of a magnetic field, the sample exhibits a spin-flop transition, which was observed in ESR, DC-field and pulsed-field magnetization measurements. 

\section{Conclusions}
Neutron powder diffraction has been used to determine the low-temperature structure of [Cu(HF$_{2}$)(pyz)$_{2}$]SbF$_{6}$. We find there are no structural phase transitions in this material from room temperature to 1.5 K. The magnetic reflections in neutron powder diffraction measurements could be indexed with the commensurate propagation vector $\textit{\textbf{k}}$ = (0, 0, 1/2) (r.l.u.). This provides experimental evidence that the weak interplane interactions in this quasi-two dimensional material are antiferromagnetic in nature, given by $J_{\perp}/k_{\textsc{b}}$ = 0.12 K. Furthermore, we found that Cu(II) moments are antiferromagnetically coupled in all directions, and within this model we have determined that the Heisenberg $S$ = 1/2 Cu(II) moments exhibit a renormalised moment of 0.6(1)$\mu_\textsc{b}$ in the ordered phase. We attribute this to the presence of quantum fluctuations present below the ordering temperature, which shows this sample a good realisation of a macroscopic quantum magnet. The reduced moment is not observed in high-field bulk magnetization measurements, emphasising that neutron powder diffraction can be used as local probe of the ordered Cu(II) moments to detect the macroscopic effects of quantum fluctuations in low dimensional coordination polymers.

We have also studied the magnetic properties of the related, isostructural, coordination polymer [Co(HF$_{2}$)(pyz)$_{2}$]SbF$_{6}$, which exhibits a transition to a long-ranged antiferromagnetically ordered state below 7.1 K. Using neutron diffraction at 4 K, we find that the magnetic peaks could be indexed with the same commensurate propagation vector as in the Cu sample, and that the magnetic ordering is G-type. For this sample, the relative intensity of the magnetic scattering implies the moments align along the $c$-axis. This shows that neutron diffraction allows for the unambiguous determination of the the nature of the single ion anisotropy of the Co(II) ions from a powdered sample.

From the heat capacity, ESR and neutron diffraction we conclude the $S$ = 3/2 Co(II) ions in this sample exhibit a ground state doublet. ESR suggests this ground state doublet is separated from the next highest doublet by at least 30~K, and this strong easy-axis anisotropy in an antiferromagnetically exchange coupled system leads to the spin-flop transition observed  in the magnetization.

The high-spin state of the Co(II) ions in conjunction with the closer plane spacing of this material permits a stronger magnetic coupling between the layers compared to the Cu system. We therefore expect that the ratio $J_{\perp}/J_{\parallel}$ is increased from the quasi-2D case of the [Cu(HF$_{2}$)(pyz)$_{2}$]SbF$_{6}$. The heat capacity data is in itself consistent with 2D or close to 3D exchange interactions, however, the absence of a broad maximum in the magnetic heat capacity suggests that the increase in $J_{\perp}$ is not so large so as to move the system to the limiting case of a 1D chain of ions coupled strongly along the $c$-axis.

This result for the exchange anisotropy in this Ising system is further consistent with the observation that the effect of quantum fluctuations are not observed below $T_{\textsc{n}}$. This is evident in that ordered magnetic moment, $\mu_{\text{Co}}$~=~3.02(6)$\mu_{\textsc{b}}$, detected from neutron diffraction at 4~K is comparable to the moment measured from ESR in the anisotropic paramagnetic phase.

By modelling the magnetic exchange with an effective spin-half model (Eq. \ref{ham}), we have shown from the magnetization that the strength of the magnetic exchange ($J$) to $N$ nearest neighbours is given by $NJ/k_{\textsc{b}}$ = 39(9)~K. Assuming the strength of the intra- and interplane interactions are of the same order ($J_{\parallel}$ = $J_{\perp}$ = $J$) we found $J/k_{\textsc{b}}$~=~7(1)~K, and the spin-exchange anisotropy $a$~=~0.28(6). The resultant value of $k_{\textsc{b}}T_{\textsc{n}}/J$~=~1.1(2) is consistent with the simple cubic model.

\begin{acknowledgments}
We thank L. Chapon for experimental assistance. JB thanks M. R. Lees for technical assistance, and S. Ghannadzadeh for the fitting function used in the heat capacity analysis. Part of this work was performed at the STFC ISIS Facility and we are grateful for the provision of beamtime. Work performed in the UK was supported by the EPSRC. The work at EWU was supported by the US NSF under grant no. DMR-1306158. A portion of this work was performed at the National High Magnetic Field Laboratory, which is supported by National Science Foundation Cooperative Agreement No. DMR-1157490, the State of Florida, and the U.S. Department of Energy.
\end{acknowledgments}

% Create the reference section using BibTeX:
\bibliography{basename of .bib file}

\end{document}